\documentclass[journal]{IEEEtran}
\IEEEoverridecommandlockouts
\usepackage{amsmath,amsfonts,amssymb, bm}
\usepackage{nccmath}
\usepackage[ruled,vlined]{algorithm2e}
\usepackage{cellspace}
\usepackage{textcomp}
\usepackage{url}
\usepackage{verbatim}
\usepackage{graphicx}
\usepackage{cite}
\usepackage[svgnames]{xcolor}
\usepackage[nolist,nohyperlinks]{acronym} 
\newcommand{\bb}[1]{\mathbb{#1}}
 \newcommand{\ten}[1]{\boldsymbol{\mathcal #1}}
 \newcommand{\ma}[1]{\boldsymbol{#1}}
 \newcommand{\ra}{\textrm{rank}}
 \usepackage{tabularx}
 \usepackage{booktabs}
 \usepackage{makecell}
 \usepackage{multirow}

 \usepackage{hyperref}
\hypersetup{hidelinks}
\def\BibTeX{{\rm B\kern-.05em{\sc i\kern-.025em b}\kern-.08em
    T\kern-.1667em\lower.7ex\hbox{E}\kern-.125emX}}
\AtBeginDocument{\definecolor{tmlcncolor}{cmyk}{0.93,0.59,0.15,0.02}\definecolor{NavyBlue}{RGB}{0,86,125}}
\newcommand{\sumx}[2]{\sum\limits_{#1}^{#2}}

\definecolor{RevThree}{RGB}{0,102,204}
\definecolor{RevFourOne}{RGB}{0,128,128}
\definecolor{RevFourTwo}{RGB}{230,126,34}
\definecolor{RevFourThree}{RGB}{128,0,128}
\definecolor{RevFourFour}{RGB}{200,0,0}
\definecolor{RevFourFive}{RGB}{0,0,255}
\definecolor{RevFourMinor}{RGB}{0,140,0}

\begin{document}
\begin{acronym}
  \acro{2G}{second generation}
\acro{3G}{third generation}
\acro{4G}{fourth generation}
\acro{5G}{fifth generation}
\acro{B5G}{beyond fifth generation}
\acro{6G}{sixth generation}
\acro{3GPP}{3$\text{rd}$~Generation Partnership Project}
\acro{LS}{least squares}
\acro{IRS}{intelligent reconfigurable surface}
\acro{RIS}{reconfigurable intelligent surface}
\acro{LIS}{large intelligent surface}
\acro{BS}{base station}
\acro{UT}{user terminal}
\acro{SU}{single-user}
\acro{MU}{multi-user}
\acro{SISO}{single-input single-output}
\acro{MISO}{multiple-input single-output}
\acro{SIMO}{single-input multiple-output}
\acro{MIMO}{multiple-input multiple-output}
\acro{CSI}{channel state information}
\acro{LOS}{line-of-sight}
\acro{NLOS}{non-line-of-sight}
\acro{SINR}{signal to interference plus noise ratio}
\acro{SNR}{signal to noise ratio}
\acro{RF}{radio frequency}
\acro{AWGN}{additive white Gaussian noise}
\acro{DFT}{discrete Fourier transform}
\acro{ALS}{alternating least squares}
\acro{TALS}{trilinear alternating least squares}
\acro{BALS}{bilinear alternating least squares}
\acro{SVD}{singular value decomposition}
\acro{HOSVD}{high order singular value decomposition}
\acro{THOSVD}{truncated high order singular value decomposition}
\acro{PARAFAC}{PARAllel FACtors}
\acro{NMSE}{normalized mean square error}
\acro{SER}{symbol error rate}
\acro{BD-RIS}{beyond-diagonal reconfigurable intelligent surface}
\acro{LS-Kron}{least squares Kronecker factorization} 
\acro{LS-KRF}{least squares Khatri-Rao Factorization}  
\acro{ZF}{zero-forcing}  
\acro{STAR-RIS}{simultaneously transmitting and reflecting reconfigurable intelligent surface}
\acro{CE}{channel estimation}
\end{acronym}


\title{Semi-Blind Joint Channel and Symbol Estimation for Beyond Diagonal Reconfigurable Surfaces}

\author{Gilderlan Tavares de Araújo, André L. F. de Almeida, Bruno Sokal, Gabor Fodor
}
\maketitle
\renewcommand\baselinestretch{.98}
\begin{abstract}
The \ac{BD-RIS} is a recent architecture in which scattering elements are interconnected to enhance the degrees of freedom for wave control, yielding performance gains over traditional single-connected RISs. For BD-RIS, channel estimation, which is well studied for conventional RIS, becomes more challenging due to complex connections and a larger number of coefficients. Prior works have relied on pilot-assisted estimation followed by data decoding. This paper introduces a semi-blind tensor-based approach to joint channel and symbol estimation that reduces the need for dedicated training sequences by directly leveraging data symbols. We consider a practical scenario with time-varying user terminal--RIS channels under mobility. By reformulating the received signal from a tensor-decomposition perspective, we develop two semi-blind receivers: a two-stage method that transforms the fourth-order PARATUCK model into a third-order PARAFAC model, and a single-stage iterative process based on the fourth-order TUCKER decomposition. Identifiability conditions for reliable joint recovery are derived, and numerical results demonstrate the performance advantages and trade-offs of the proposed schemes over existing solutions.
\end{abstract}



\section{INTRODUCTION}
\IEEEPARstart{R}{econfigurable} surfaces are a promising technology for future wireless communication systems \cite{RIS_6G_2021,RIS_Application,RIS_Opotunit}. Most prior attention has been devoted to the passive \acf{RIS} architecture, which is equipped solely with reflection capabilities. Recently, new architectures have been intensively studied to address the limitations of traditional \ac{RIS} and improve system performance. Examples include hybrid \ac{RIS} architectures \cite{Hibrid_Alexandropoulos_2023_MAG,Hibrid_Alexandropoulos_2023_TCOM,Amarilton_2024_ISWCS} that mitigate the double fading effect and the \ac{STAR-RIS} architecture \cite{STAR_2023_JSTSP,STAR_2023_TCOM,STAR_RIS_360_mag}, which supports reflection and transmission capabilities.

A new \ac{RIS} architecture called beyond-diagonal \ac{RIS} was recently proposed in \cite{Clerckx_TWC_APR_2023, B_Clerckc_CE_TSP_24}. The so-called \acf{BD-RIS} introduces additional degrees of freedom for system optimization, resulting in significant performance improvements compared to the conventional diagonal \ac{RIS} (single-connected, i.e., there are no connections between the \ac{RIS} elements). The non-diagonal structure of the phase shift matrix also emerges in scenarios involving non-reciprocal connections among \ac{RIS} elements (e.g., group-connected and fully-connected cases), allowing an incident signal to be reflected by a different element \cite{Clerckx_TVT_JUN_2022}. In \cite{Shen_2022}, a modeling and architecture design based on scattering network analysis was proposed. More recently, \cite{matteo_2024} introduced a \ac{RIS} modeling framework grounded in multiport network analysis, while optimization algorithms for BD-\ac{RIS} were explored in \cite{Clerckx_TWC_FEB_2024, B.Clerckx_TWC_EA_2024}. Despite its potential, this technology faces several challenges, including channel estimation, joint optimization of passive and active beamforming, and mitigation of multi-user interference \cite{zheng_survey}. 

The performance enhancements promised by BD-\ac{RIS} relative to traditional single-connected \ac{RIS} depend strongly on the accuracy of \ac{CSI}. In channel estimation, most efforts have focused on traditional single-connected \ac{RIS} architectures, in which the phase-shift matrix is assumed to be diagonal, implying no physical coupling among \ac{RIS} elements. However, the lack of signal-processing capabilities in BD-\ac{RIS}, combined with its more complex connection structure, makes channel estimation even more challenging than in a single-connected \ac{RIS}. Few solutions have been proposed to estimate the channel in BD-\ac{RIS}. In \cite{B_Clerckc_CE_TSP_24}, a \ac{LS} method was introduced to estimate the cascaded channel (\ac{BS} to \ac{BD-RIS} to \acf{UT} channel). Despite its conceptual simplicity, this method requires significant training overhead. The work \cite{Sokal_BD_RIS_2024} proposed closed-form and iterative algorithms to estimate the BD-\ac{RIS} channel that leverage tensor decomposition techniques. Although the approach of \cite{B_Clerckc_CE_TSP_24} only estimates the cascaded channel, the tensor-based methods proposed in \cite{Sokal_BD_RIS_2024} provide individual estimates of the channel matrices, thereby improving performance while significantly reducing training overhead. Nevertheless, these methods are based on pilot sequences, implying that CSI acquisition and data detection are performed in two consecutive stages. More recently, \cite{Gil_Asilomar} proposed a semi-blind joint channel and symbol estimation approach for BD-\ac{RIS}. In an \ac{MIMO} communication scenario, this method directly produces decoupled estimates of the involved channel matrices using a PARATUCK decomposition of the received signals \cite{Gil_TSP}. A framework for channel estimation in both RIS and BD-RIS scenarios is proposed in \cite{Suggestion_RI}. Another related line of work relies on received power measurements for channel estimation. Such methods have been proposed for both conventional RIS/IRS and BD-RIS systems to infer channel information while reducing the dependence on explicit pilot signaling \cite{Sun2024PowerMeasurementIRS,Sun2025PowerAutocorrelationIRS,Liu2025PowerMeasurementBDRIS}. Unlike the proposed semi-blind tensor-based framework, these approaches follow a different estimation paradigm that relies on power observations rather than on a tensor modeling of the received signals.

Tensor decompositions are powerful mathematical tools for modeling and estimating the multidimensional structure of signals and data. As demonstrated by the extensive signal processing literature, tensor modeling provides a concise and elegant framework for addressing various problems, such as blind/semi-blind channel estimation and data detection in wireless communications \cite{Almeida_Elsevier_2007,Almeida2008,Favier2014,Ximenes_2015,Ximenes_2014,Chen2021}. More recently, tensor-based receiver algorithms have proven to be highly effective in RIS-assisted MIMO communication systems \cite{Fazal_2024_TVT,Fazal_2024_TCOM}. In the context of single-connected RIS-assisted communications, tensor modeling approaches have been exploited to solve problems such as phase shift optimization \cite{Yuri_Sales_2024_WCL}, feedback overhead reduction \cite{Sokal_2023_TWC}, and channel estimation \cite{Gil_JTSP,paulo2022tensor,Gherekhloo23,Nwalozie24}. To our knowledge, semi-blind joint channel and symbol estimation for \ac{BD-RIS} has not yet been well investigated in the literature. Leveraging data symbols for channel estimation avoids transmitting lengthy pilot sequences, while using information-bearing symbols to refine channel estimates. This approach also reduces decoding latency, since early detection can be performed jointly with channel estimation. 

This paper considers semi-blind joint channel and symbol estimation methods for \ac{BD-RIS}. The proposed data-driven solutions produce decoupled estimates of the communication channels involved (\ac{UT} to \ac{BD-RIS} and \ac{BD-RIS} to \ac{BS}) without a dedicated prior training-sequence stage. This can improve spectral efficiency and reduce data-decoding latency compared to pilot-assisted methods, since useful data-symbol detection already occurs during the channel-estimation stage. 

Adopting a novel non-diagonal scattering design for semi-blind channel estimation, where the BD-\ac{RIS} phase shift matrix is expressed as the product of a scattering matrix and a diagonal rotation matrix, the received signal can be recast as a fourth-order tensor that follows a PARATUCK-based model. Using the resulting algebraic structure of the signal, we derive two novel receiver algorithms for joint channel and symbol estimation. The first recasts the received signal tensor as a structured third-order PARAFAC model, from which channel and symbol estimates are obtained in two stages. In the second case, the received signal is reshaped into a fourth-order TUCKER tensor, enabling us to derive a single-stage channel and symbol estimation scheme based on the \ac{TALS} algorithm. Our method is considered ”semi-blind” because it jointly estimates the channel and symbols without relying on a dedicated pilot training period, unlike pilot-assisted techniques.\footnote{The proposed method is designated as semi-blind rather than fully blind because a small set of identification symbols is assumed to be known at the receiver.} Compared with the conference version in \cite{Gil_Asilomar}, this paper provides a complete formulation of the proposed semi-blind \ac{BD-RIS} channel-estimation problem, presents detailed derivations of the two proposed algorithms, and discusses identifiability conditions and their implications for system design while accounting for channels that vary over different time scales due to user mobility. Extensive numerical results are also provided to evaluate performance across various setups. 

The main contributions of this work are as follows.
\begin{itemize}
    \item We propose a data-driven channel estimation approach for \ac{BD-RIS}, where separate estimates of the involved channel matrices are obtained without pilot sequences by exploiting a tensor algebraic signal structure based on a PARATUCK-(2,4) decomposition.
    \item By exploiting two different tensor reshapings of the received signal, we derive two semi-blind receivers. We first formulate a two-stage receiver that yields estimates of the two channel matrices and the symbol matrix in two consecutive stages by recasting the received signal as a structured PARAFAC model. Then, a second receiver is derived that iteratively solves the problems of channel and symbol estimation in a single stage by resorting to fourth-order TUCKER modeling. 
    \item We study the identifiability conditions for the joint channel and symbol estimation, considering the two proposed receivers. We also discuss the design of the \ac{BD-RIS} scattering matrix and the computational complexity of the proposed methods. Our study sheds light on the trade-offs between the two proposed receivers, accounting for both complexity and performance. 
    \item  Extensive numerical results are provided to evaluate the performance of the proposed semi-blind receivers. Our results show that both algorithms perform well, although the TUCKER receiver is more efficient than PAKRON\footnote{This acronym is a combination of PARAFAC and the Kronecker receiver} while being computationally more complex. Additionally, we compare our proposed method with the pilot-assisted method, which performs better than traditional LS and worse than other tensor methods for composite channel estimation. However, our method is semi-blind, which improves spectral efficiency compared with both competing methods.
\end{itemize}

The remainder of the paper is organized as follows. Section II presents the system model, while Section III details the two proposed receivers. In Section IV, we discuss identifiability and complexity, as well as the trade-offs associated with the proposed receivers. The simulation results are provided in Section V, while our concluding remarks are given in Section VI.\footnote{Although this is outside the scope of the current study, it is important to consider BD-RIS scenarios involving hardware imperfections in future research, as has been done in other studies \cite{stacked,holographic,Performance}.}


\section{TENSOR PRELIMINARIES}

In this section, we introduce the notation and main operators used in this paper and provide a brief description of the TUCKER-4 and PARATUCK-(2,4) decompositions, which will be used to develop the proposed semi-blind receivers. In particular, we provide the formulation of these decompositions in scalar and frontal-slice notations.

\subsection{Notation and properties}
Matrices are represented with bold capital letters $(\ma{A})$, and vectors are denoted by bold lowercase letters $(\ma{a})$. Tensors are symbolized by calligraphic letters $(\ten{A})$. Hermitian, Transpose and pseudo-inverse of a matrix $\ma{A}$ are denoted as $\ma{A}^{\text{H}}$, $\ma{A}^{\text{T}}$ and $\ma{A}^\dagger$, respectively.  $\ma{A}_{i.}$ and $\ma{A}_{. j}$ denote the $i$-th row and $j$-th column of the matrix $\ma{A}$, respectively. $\ma{D}_i(\ma{A}) = \text{diag}(\ma{A}_{i.})$ is a diagonal matrix that holds the $i$-th row of $\ma{A}$ on its main diagonal, since the operator $\text{diag}(\ma{a})$ forms a diagonal matrix out of its vector argument.  $\ast$, $\circ$, $\diamond$, $\odot$, and $\otimes$ denote conjugate, outer product, Khatri Rao, Hadamard, and Kronecker products, respectively. $\ma{I}_N$ denotes the $N \times N$ identity matrix. The operator $\text{vec}(\cdot)$ vectorizes an $I \times J$ matrix argument, while $\text{unvec}_{I \times J}(\cdot)$ does the opposite operation. Moreover, $\text{vecd}(.)$ forms a vector out of the diagonal of its matrix argument. In this paper, we make use of the following identities:
\begin{equation}
\textrm{vec}(\ma{ABC}) = (\ma{C}^{\textrm{T}} \otimes \ma{A})\textrm{vec}(\ma{B}).
\label{Eq:Propertie Vec General}
\end{equation}
\begin{equation}
\textrm{diag}(\ma{a})\ma{b} = \textrm{diag}(\ma{b})\ma{a}.
\label{Eq:Propertie diag(a)b}
\end{equation}
If $\ma{B}$ is a diagonal matrix, we have:
\begin{equation}
\textrm{vec}(\ma{ABC}) = (\ma{C}^{\text{T}} \diamond \ma{A})\text{vecd}(\ma{B}).
\label{Eq:Propertie Vec restrict}
\end{equation}
\begin{equation}
    (\ma{A} \otimes \ma{B})(\ma{C} \otimes \ma{D)} = \ma{AC} \otimes \ma{BD}
    \label{prop: Kron_product}
\end{equation}
\begin{equation}
    (\ma{A} \otimes \ma{B})^{\text{H}} = \ma{H}^{\text{H}} \otimes \ma{D}^{\text{H}}
    \label{prop: Kron_Inverse}
\end{equation}

The $n$-mode product between a tensor $\ten{Y} \in \bb{C}^{I_1 \times \ldots I_n \ldots \times I_N}$ and a matrix $\ma{A} \in \bb{C}^{I_n \times R}$ is denoted as $\ten{Y}\times_n\ma{A}$, for $1 \leq n \leq N$. An identity $N$-way tensor of dimension $R\times R \cdots \times R$ is denoted as $\ten{I}_{N,R}$. $\ma{1}_{M,N}$ is an $M \times N$ one matrix. $\sqcup_n$ denotes the concatenating operator related to the $n$-mode. The tensor $\ten{Y}$ can be \textit{matricized} by letting one dimension vary along the rows and the remaining two dimensions along the columns. From $\ten{Y} \in \bb{C}^{I_1 \times \ldots I_n \ldots \times I_N}$, we have $N$ different matricizations, each one referred to as the \textit{$n$-mode unfolding}, $n=1,2,\ldots, N$. We refer the interested reader to \cite{Sidiropoulos2017} for an overview.

\subsection{TUCKER-4 decomposition}

The Tucker decomposition \cite{Kolda2009} defines the concept of multilinear transformation. For a fourth-order tensor $\ten{Y} \in \bb{C}^{I \times J \times K \times L}$, it expresses the tensor as multiple sums of rank-one tensor components, which can be defined as 
 \begin{align}
\ten{Y} &= \sumx{p=1}{P} \sumx{q=1}{Q} \sumx{r=1}{R}\sumx{s=1}{S}g_{p,q,r,s}a_{i,p} b_{j,q}c_{k,r}d_{l,s}
,\label{eq:tucker_model} 
\end{align}
where $\ma{A} \in \bb{C}^{I \times P}$, $\ma{B} \in \bb{C}^{J \times Q}$, and $\ma{C} \in \bb{C}^{K \times R}$, and $\ma{D}  \in \bb{C}^{L \times S}$ are the factor matrices, and $\ten{G} \in \bb{C}^{P \times Q\times R\times S}$ is referred to as the \emph{core tensor}, with typical elements given by $a_{i,p}, b_{j,q}, c_{k,r}, d_{l,s}$, and $g_{p,q,r,s}$, respectively.  
Adopting the $n$-mode product notation, the Tucker-4 decomposition can be written as
\begin{align}
  \label{eq:tenZ_tucker_generic} \ten{Y} &= \ten{G} \times_1 \ma{A} \times_2 \ma{B} \times_3 \ma{C} \times_4 \ma{D}
\end{align}
Note that when $L=S=1$, it reduces to a Tucker-3 decomposition of a third-order tensor  $\ten{Y} \in \bb{C}^{I \times J \times K}$.

\subsection{PARATUCK-(2,4) decomposition}

A generalized PARATUCK-type decomposition for tensors was proposed in \cite{favier2012tensor, de2013space} as an extension to the PARATUCK-2 decomposition to fourth-order tensors, also known as PARATUCK-(2,4) tensor decomposition. Its scalar form is given by:
\begin{equation}\label{EQ:scalar notation of PT(2-4)}
x_{i,j,k,l} = \sum_{p = 1}^{P}\sum_{q = 1}^{Q}a_{i,p}b_{j,q}c_{k,p}^{\ma{A}}c_{k,q}^{\ma{B}}g_{p,q,l}\;,
\end{equation}
where $\ma{A} \in \bb{C}^{I \times P}$, $\ma{B} \in \bb{C}^{J \times Q}$ are the factor matrices, $\ma{C}^{\mathbf{A}} \in \bb{C}^{K \times P}$, $\ma{C}^{\mathbf{B}} \in \bb{C}^{K \times Q}$ are referred to as interactions matrices, and $\ten{G} \in \bb{C}^{P \times Q \times L}$ is the \emph{core tensor}.

A useful notation to express the PARATUCK-(2,4) decomposition, consisting of the most common matrix-slice notation, obtained by fixing the indices $(k,l)$, is given by:
\begin{equation}
[\ten{X}]_{..kl} = \ma{A}\ma{D}_{k}(\ma{C}^{\ma{A}})\ma{G}_{..l}\ma{D}_{k}(\ma{C}^{\ma{B}})\ma{B}^{\text{T}} \; ,
\label{EQ: Slice PARATUCK24}
\end{equation}
where $\ma{G}_{...l}$ is the $l$th frontal slice of the tensor $\ten{G} \in \bb{C}^{P \times Q \times L}$. Note that PARATUCK-2 (2,4) reduces to PARATUCK-2 (also referred to as PARATUCK-(2,3) \cite{Favier2014}) if $L=1$, i.e., when the core tensor $\mathcal{G} \in \mathbb{C}^{P \times Q \times L}$ simplifies to a core matrix $\ma{G} \in \bb{C}^{P \times Q}$.

\section{SYSTEM MODEL}
\begin{figure}[!t]
    \centering
    \includegraphics[scale = 0.42]{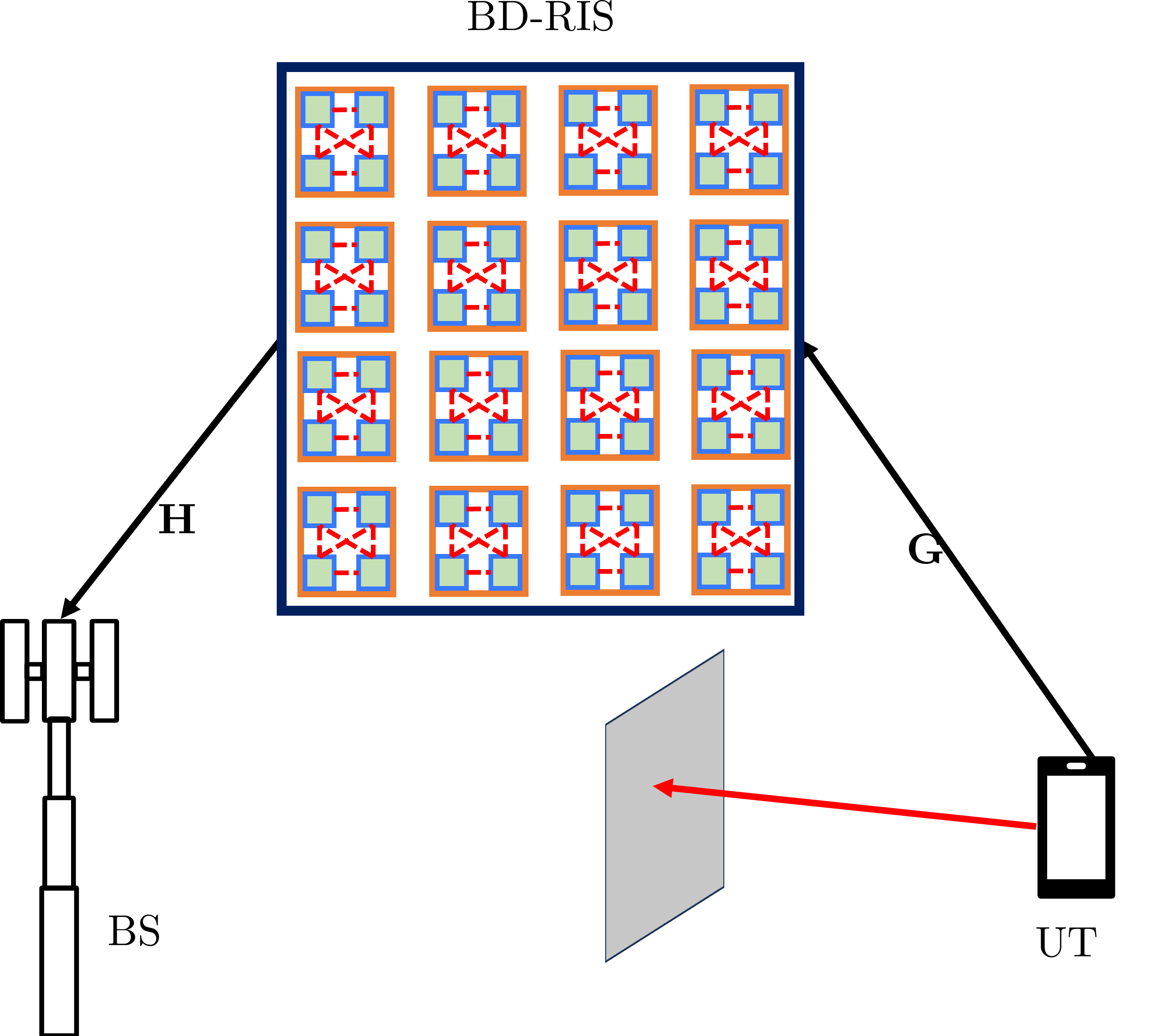}
    \caption{Uplink BD-RIS assisted MIMO system}
    \label{fig: system model}
\end{figure}

Consider a single-user narrowband \ac{BD-RIS}-assisted uplink MIMO system, as illustrated in Fig.~\ref{fig: system model}, where the transmitter and receiver are equipped with $M_T$ and $M_R$ antennas, respectively, and the \ac{BD-RIS} has $N$ elements. Different schematic BD-RIS architectures, including conventional RIS, can be found in \cite{Clerckx_TWC_APR_2023}. The direct link between the transmitter and the receiver is assumed to be very weak or unavailable.\footnote{If the direct channel is available, it can be estimated using a two-step approach similar to the conventional RIS case \cite{Gil_TSP}. First, the BD-RIS is switched OFF, and the direct channel is estimated. Then it is switched ON, and the contribution of the direct link is subtracted, enabling identification of the assisted-link channel.} We consider a group-connected \ac{BD-RIS} architecture with $Q$ groups, comprising $\overline{N} = \frac{N}{Q}$ elements each, where the elements within each group are all connected. Additionally, residual inter-group leakage is considered to be sufficiently small to be neglected or treated as part of the additive noise term \cite{Clerck_Tutorial}.  The signal received in the $t$-th symbol slot can be expressed as \cite{B_Clerckc_CE_TSP_24,Clerckx_CAMSAP_2023}
\begin{equation}
    \ma{y}_{t} = \sum_{q = 1}^{Q}\ma{H}^{(q)}\ma{S}^{(q)}\ma{G}^{(q)}\ma{x}_{t} + \ma{z}_{t}^{(q)},
    \label{Eq: Rx at (k,t) without KR code}   
\end{equation}
where $\ma{H}^{(q)} \in \bb{C}^{M_R \times \overline{N}}$ and $\ma{G}^{(q)} \in \bb{C}^{\overline{N} \times M_T}$ represent the \ac{RIS}-BS and UT-\ac{RIS} channels, respectively, while $\ma{S}^{(q)} \in \bb{C}^{\overline{N} \times \overline{N}}$ denotes the (non-diagonal) scattering matrix associated with the $q$-th group. Additionally, $\ma{x}_{t} \in \bb{C}^{M_T \times 1}$ represents the data symbol vector, while $\ma{z}_{t}^{(q)}$ corresponds to the additive white Gaussian noise (AWGN) term.

\begin{figure}[!t]
    \centering
    \includegraphics[scale = 1.0]{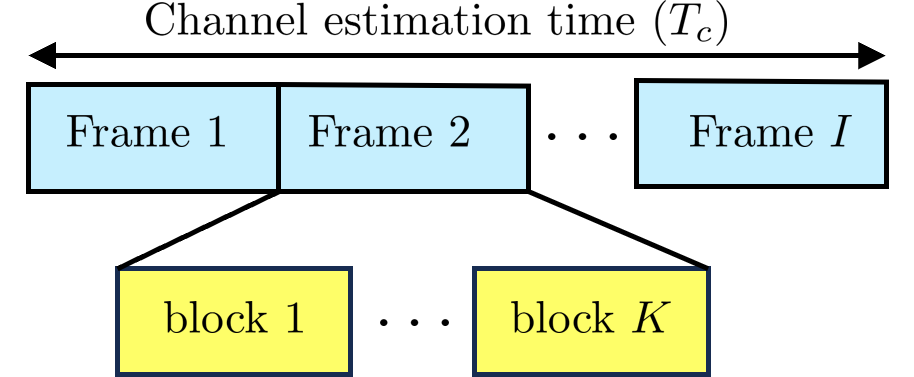}
    \caption{Transmission protocol time structure: The channel estimation slot is divided into $I$ frames, each frame composed of $K$ blocks.}
    \label{fig: Time vary protocol}
\end{figure}

Considering the transmission protocol illustrated in Figure \ref{fig: Time vary protocol}, we consider a structured block transmission scheme, where the semi-blind \ac{CE} period consists of $I$ frames, where each frame contains $K$ data blocks of $T$ symbol slots each, such that $T_c = KTI$. For each frame, the scattering matrix is assumed to change from block to block, while remaining unchanged during a block \cite{Sokal_Asilomar_2024}.  
We assume that the data symbols in each block are encoded using a Khatri-Rao coding scheme\footnote{The Khatri-Rao coding used here induces the structured tensor model exploited by the proposed receivers. This structure enables decoupled estimation of the involved channels, i.e., the separate recovery of the BD-RIS-BS and UT-BD-RIS channel matrices, rather than estimating only a composite cascaded channel.} \cite{Gil_Asilomar}. Moreover, to account for some user mobility, the \ac{UT}-\ac{RIS} channel linking the user to the RIS is assumed to vary from frame to frame. In contrast, the RIS-BS channel remains constant throughout the transmission interval. This assumption is reasonable because the \ac{BS} and \ac{RIS} are typically deployed at fixed locations, whereas the UT is a mobile node.\footnote{ In high-mobility conditions where the UT--BD-RIS channel varies very quickly (for instance, at each symbol interval), the specific tensor model introduced in this paper becomes restrictive. In such situations, the tensor-based framework can still be employed, but it must be adapted to use alternative tensor modeling methods that more effectively track rapid channel fluctuations \cite{Kenneth_WCL}.} Under these assumptions, the signal received during the $t$-th symbol slot of the $k$-th block and $i$-th frame can be expressed as
\begin{equation}
    \ma{y}_{i,k,t} = \sum_{q = 1}^{Q}\ma{H}^{(q)}\ma{S}_{k}^{(q)}\ma{G}^{(q)}_i\text{diag}(\ma{w}_k)\ma{x}_{t} + \ma{z}_{i,k,t}^{(q)},
    \label{Eq: Rx at (i,k,t) with KR code}
\end{equation}
where $\ma{w}_k$ is the coding vector in the $k$-th block. Collecting the signals received during the $T$ time intervals in a given frame $i$ and block $k$, we get the following.
\begin{equation}
\begin{aligned}
    \ma{Y}_{i,k} & = \sum\limits_{q=1}^Q\boldsymbol{H}^{(q)}\ma{S}_{k}^{(q)}\ma{G}^{(q)}_i\text{diag}(\ma{w}_{k})\ma{X}^{\text{T}} + \ma{Z}_{i,k}\\
    & = \Big(\sum\limits_{q=1}^Q\boldsymbol{H}^{(q)}\ma{S}_{k}^{(q)}\ma{G}^{(q)}_i\Big)\text{diag}(\ma{w}_{k})\ma{X}^{\text{T}} + \ma{Z}_{i,k} \in \bb{C}^{M_r \times T},
\end{aligned}
    \label{Eq: Block DB_RIS_PT2}
 \end{equation}
 where $\ma{X} = \left[\ma{x}_1, \ldots , \ma{x}_T\right]^{\text{T}} \in \bb{C}^{T \times M_T}$ denotes the data symbol matrix and $\ma{Z}_{i,k} \in \bb{C}^{M_r \times T}$ is the AWGN term. The proposed framework assumes a multi-timescale block-fading model, in which the BS-RIS channel remains approximately constant during the estimation interval, whereas the UT-RIS channel may vary from frame to frame due to user mobility.

\vspace{-0.5cm}

\subsection{SCATTERING MATRIX DESIGN}
\label{Sec: BD-RIS Design}
The design of the \ac{BD-RIS} scattering matrix used for training is fundamental to the uniqueness of the separate estimates of the communication channels involved. A baseline design for an \ac{LS} channel estimation was first proposed in \cite{B_Clerckc_CE_TSP_24,Clerckx_CAMSAP_2023}, accounting for the physical constraints of the \ac{BD-RIS} while ensuring energy conservation and the unitary property. Here, we consider an alternative scattering-matrix design tailored to the proposed semi-blind channel and symbol-estimation approach. Starting from the fundamental signal model in (\ref{Eq: Block DB_RIS_PT2}), we propose designing the BD-\ac{RIS} matrix as the product of a fixed (static) scattering matrix $\ma{S}_0$ and a time-varying block-dependent diagonal rotation such that 
\begin{equation}\label{eq:smd}
\ma{S}_k^{(q)} = \ma{S}_0\text{diag}(\ma{\overline{p}}_{k}^{(q)}),
\end{equation}
where $\ma{\overline{p}}_{k}^{(q)}$ is a vector containing the phase rotation angles associated with the $q$-th group in the $k$-th block. Unlike the state-of-the-art \cite{B_Clerckc_CE_TSP_24} and \cite{Sokal_BD_RIS_2024,Sokal_Asilomar_2024}, the proposed design has a lower complex implementation since only the diagonal term $\ma{\overline{p}}_{k}^{(q)}$ changes per group. The proposed design also maintains the unitary property of the scattering matrix, as in \cite{B_Clerckc_CE_TSP_24}, i.e.,
\begin{equation}
(\ma{S}_{k}^{(q)})^{\text{H}}\ma{S}_{k}^{(q)} = \ma{S}_{k}^{(q)}(\ma{S}_{k}^{(q)})^{\text{H}} = \ma{I}_{\overline{N}} \; \; \; \forall q. 
\label{EQ: identity condition for S_k}
\end{equation} Note that constraint (\ref{EQ: identity condition for S_k}) stems from the assumption of a lossless multiport network model, following the theory in \cite{Clerckx_TWC_APR_2023,Clerck_Tutorial}. Since $\ma{S}_k^{(q)} = \ma{S}_0\text{diag}(\ma{\overline{p}}_{k}^{(q)})$, $q=1,\ldots, Q$, this constraint holds if $\ma{S}_0^{\text{H}}\ma{S}_0 = \ma{S}_0\ma{S}_0^{\text{H}} = \ma{I}_{\overline{N}}$ and $\ma{\overline{p}}_{k}^{(q)}$ is a vector of complex exponentials.

Note that in this work, we take another step towards simplifying the BD-RIS training design. Instead of changing the full BD-RIS scattering matrix $\ma{S}^{(q)}_k$ in each time block, $q=1,\ldots, Q$, which implies changing the scattering coefficients $KQ\bar{N}^2$ throughout training, we only change the ``diagonal'' component that defines $\ma{S}_k^{(q)} = \ma{S}_0\text{diag}(\ma{\overline{p}}_{k}^{(q)})$, $q=1,\ldots, Q$. 
Hence, the proposed BD-RIS training design entails only changing the $KQ\bar{N}$ phase values during training. Note that, for larger values of $\bar{N}$, i.e., for strongly connected architectures, we can achieve a significant reduction in the number of adjustable BD-RIS parameters. From a practical implementation perspective, dynamically reconfiguring inter-element connections (as in \cite{Sokal_BD_RIS_2024} and \cite{Sokal_Asilomar_2024}) entails greater hardware complexity than simply tuning the element phases, as proposed in this work. This simplified design may not be globally optimal for maximizing spectral or energy efficiency, since it is selected to satisfy the identifiability requirements of the proposed semi-blind estimator. Hence, a trade-off exists between training/estimation identifiability and communication-oriented beamforming optimality. In the numerical comparisons, the proposed receivers use the design in (\ref{eq:smd}), whereas the pilot-assisted baselines are evaluated with their original training/scattering designs from the corresponding references. Unless otherwise stated, the entries of the phase-rotation vectors are generated as
$p_k(n)=e^{j\phi_{k,n}}$, where $\phi_{k,n}$ are independently drawn from a uniform distribution over $[0,2\pi)$.

\subsection{RECEIVED SIGNAL MODEL}    
\begin{figure}[h!t]
  \centering
  \includegraphics[scale = 0.34]{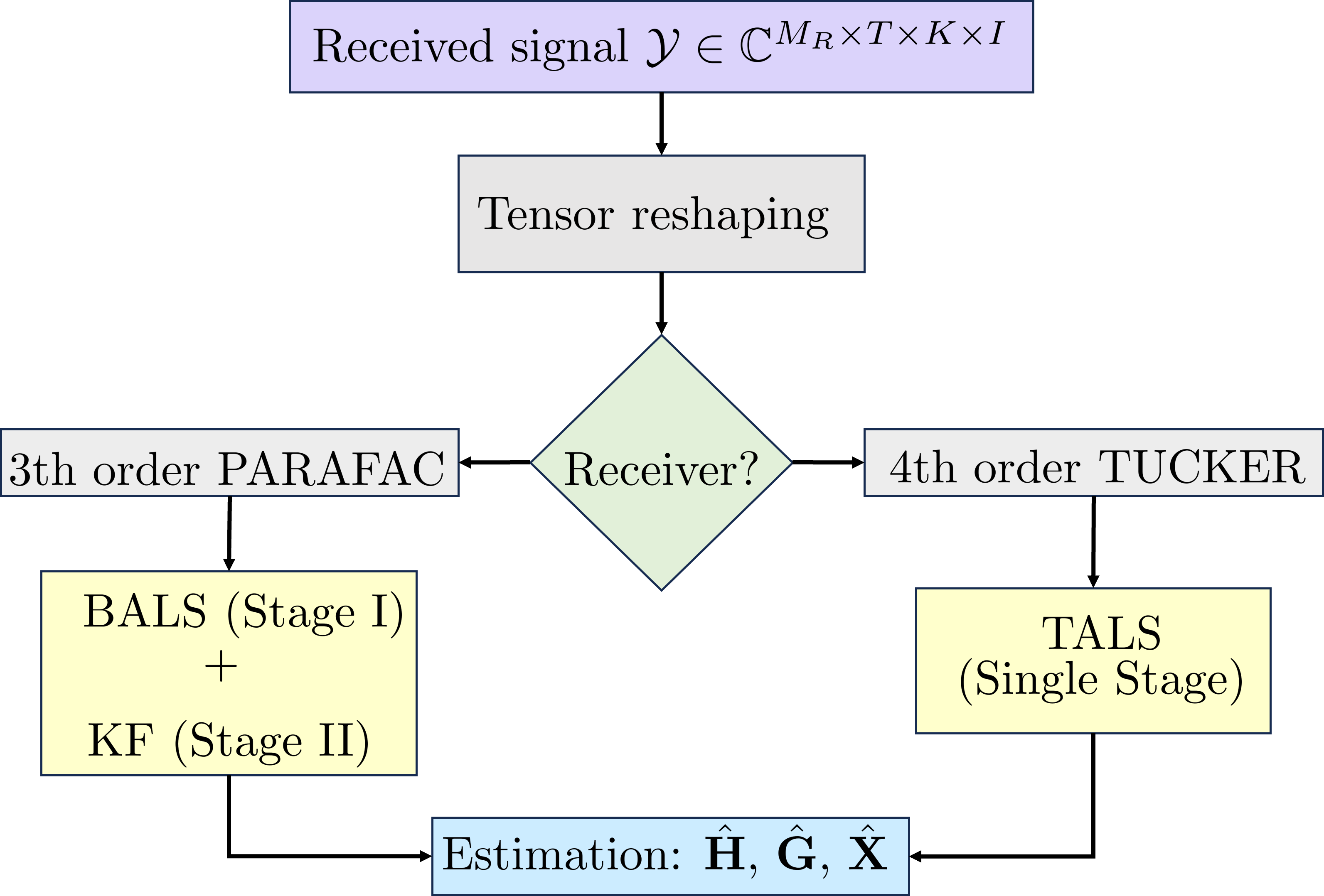}
      \caption{Flowchart summarizing the proposed receivers.}
      \label{fig:fluxograma}
  \end{figure}

Considering the scattering matrix design discussed in Section \ref{Sec: BD-RIS Design}, the noiseless$\footnote{We omit the noise term here for notation convenience. The noise contribution will be added later.}$ received signal given in (\ref{Eq: Block DB_RIS_PT2}) associated with frame $i$ and block $k$ can be expressed as
\begin{equation}
\begin{aligned}
    \ma{Y}_{k,i} &  = \Big(\sum_{q = 1}^{Q}\ma{H}^{(q)}\ma{S}_{0}\text{diag}(\ma{\overline{p}}_{k}^{(q)})\ma{G}^{(q)}_i\Big)\text{diag}(\ma{w}_{k})\ma{X}^{\text{T}},  
\end{aligned}
\label{Eq:Rx siganl at (k,t)}
\end{equation}
which can be equivalently written as
\begin{equation}
\begin{aligned}
    \ma{Y}_{k,i} &  = \Big[\ma{H}^{(1)}\ma{S}_{0} \dots \ma{H}^{(Q)}\ma{S}_{0}\Big]\text{diag}(\ma{p}_{k})\ma{G}_i\text{diag}(\ma{w}_{k})\ma{X}^{\text{T}}  \\
     & =  \Big[\ma{H}^{(1)} \dots \ma{H}^{(Q)}\Big]\ma{S}\text{diag}(\ma{p}_{k})\ma{G}_i\text{diag}(\ma{w}_{k})\ma{X}^{\text{T}} \\
     & = \ma{H}\ma{S}\text{diag}(\ma{p}_{k})\ma{G}_i\text{diag}(\ma{w}_{k})\ma{X}^{\text{T}}
     \end{aligned}
     \end{equation}
     or, compactly, 
     \begin{equation}
       \ma{Y}_{i,k} =   \ma{H}\ma{S}\ma{D}_k(\ma{P})\ma{G}_i\ma{D}_k(\ma{W})\ma{X}^{\text{T}},
        \label{Eq:Fundamental Equation}
\end{equation}
where $\ma{G}_i = \left[(\ma{G}_i^{(1)})^{\text{T}} \;  \dots \;
  (\ma{G}_i^{(Q)})^{\text{T}}\right]^{\text{T}} \in \bb{C}^{N \times M_T}$ and $\ma{H} = \left[\ma{H}^{(1)} \;  \dots \;
  \ma{H}^{(Q)}\right] \in \bb{C}^{M_R \times N}$ are the two involved channel matrices that collect the contributions of all groups.  Let us define $\ma{p}_k\doteq[\ma{p}^{(1)\text{T}}_k, \ldots, \ma{p}^{(Q)\text{T}}_k]^{\text{T}} \in \bb{C}^{N \times 1}$ and $\ma{S} \doteq \text{bdiag}(\ma{S}_{0}, \ldots \ma{S}_{0}) \in \bb{C}^{N \times N}$. Additionally, define $\ma{P}=[\ma{p}_1, \ldots, \ma{p}_K]^{\text{T}} \in \bb{C}^{K \times N}$ and $\ma{W}\doteq [\ma{w}_1, \ldots, \ma{w}_K]^{\text{T}} \in \bb{C}^{K \times M_T}$ as the BD-\ac{RIS} phase rotation matrix and the transmit coding matrix, respectively. Comparing with (\ref{EQ: Slice PARATUCK24}), the noiseless signal model in (\ref{Eq:Fundamental Equation}) corresponds to a PARATUCK-(2,4) decomposition of the fourth-order tensor $\ten{Y} \in \bb{C}^{M_R \times T \times K \times I}$, which belongs to the family of PARATUCK-type decompositions \cite{Andre_EURASIP_2014}. As a hybrid of PARAFAC and TUCKER decompositions, it combines the uniqueness of the former with the flexibility of the latter. Based on the fundamental equation in (\ref{Eq:Fundamental Equation}), we derive two receivers, which are summarized in Figure \ref{fig:fluxograma}

\section{SEMI-BLIND RECEIVERS}
We propose a data-driven approach to obtain separate estimates of the involved channel matrices without requiring pilot sequences. This can be achieved by exploiting a tensor algebraic signal structure according to a PARATUCK-(2,4) decomposition of the received signal in (\ref{Eq:Fundamental Equation}). We formulate two semi-blind receivers that exploit the intrinsic algebraic structure of the received signal tensor in two different ways. The common feature of both receivers is to leverage useful data symbols to achieve decoupled estimates of the two channel matrices. The tradeoffs between the two receiver algorithms in terms of performance, complexity, and operating conditions are also discussed.

\subsection{PARAFAC-Based Receiver}
\label{Sec: PARAFAC design}
The received signal model in (\ref{Eq:Fundamental Equation}) involves three unknown system matrices, namely the static channel matrix $\ma{H}$ linking the \ac{BD-RIS} and the BS, the block-dependent channel matrix $\ma{G}_i$ linking the UT and \ac{BD-RIS}, and the symbol matrix $\ma{X}$. Note that the vectorized version of the received signal in frame-$i$ block-$k$ is
\begin{equation}
    \ma{y}_{i,k} = \text{vec}(\ma{Y}_{i,k}) = \Big(\ma{X} \otimes \ma{HS}\Big)\Big(\ma{D}_k(\ma{W}) \otimes \ma{D}_k(\ma{P})\Big)\ma{g}_i,
    \label{EQ: vec form of Y_{i,k}}
\end{equation}
where $\ma{g}_i = \text{vec}(\ma{G}_i) \in \bb{C}^{M_TN}$. Applying property (\ref{Eq:Propertie diag(a)b}) in (\ref{EQ: vec form of Y_{i,k}}), we have
\begin{equation}
\begin{aligned}
\text{vec}(\ma{Y}_{i,k})  = \Big(\ma{X} \otimes \ma{HS}\Big)\text{diag}(\ma{g}_i)\Big(\ma{W}_{k\cdot}^{\text{T}} \otimes \ma{P}_{k\cdot}^{\text{T}}\Big)\;. 
\end{aligned}
\end{equation}
Then, collecting the signal from all the $K$ data blocks $\ma{Y}_i = \Big[\text{vec}(\ma{Y}_{i,1}) \; \text{vec}(\ma{Y}_{i,2}) \; \dots \; \text{vec}(\ma{Y}_{i,K})\Big]$, we have the signal received in frame $i$  as 
\begin{equation}
\begin{aligned}
\ma{Y}_{i} & = \Big[\ma{y}_{i,1} \; \ma{y}_{i,2} \; \dots \; \ma{y}_{i,K}\Big]\\
            & = \Big(\ma{X} \otimes \ma{HS}\Big)\text{diag}(\ma{g}_i)\Big[\ma{W}_{1\cdot}^{\text{T}} \otimes \ma{P}_{1\cdot}^{\text{T}}\; \dots \; \ma{W}_{K\cdot}^{\text{T}} \otimes \ma{P}_{K\cdot}^{\text{T}}\Big]\;\\
            & = \Big(\ma{X} \otimes \ma{HS}\Big)\text{diag}(\ma{g}_i)\Big(\ma{W}^{\text{T}} \diamond \ma{P}^{\text{T}}\Big)\; \in \bb{C}^{M_RT \times K}.
\end{aligned}
\label{EQ: Y_i}
\end{equation}
Indeed, (\ref{EQ: Y_i}), is a $3$-mode unfolding transposed $[\ten{Y}_i]_{(3)}^\text{T}$ of the tensor $\ten{Y}_i \in \bb{C}^{M_R \times T \times K}$ that fits a third-order TUCKER tensor model whose $n$-mode product is given as 
\begin{equation}
    \ten{Y}_i  = \ten{T}_{\ma{G}_i}
 \times_1 \ma{HS} \times_2 \ma{X} \times_3 \Big(\ma{W}^{\text{T}} \diamond \ma{P}^{\text{T}}\Big)^\text{T},
 \end{equation}
 where $\ten{T}_{\ma{G}_i} \in \bb{C}^{N \times M_T \times M_TN}$ such that the $3$-mode unfolding of the tensor $[\ten{T}_{\ma{G}_i}]_{(3)}$ coincides with $\text{diag}(\ma{g}_i)$, that is, $[\ten{T}_{\ma{G}_i}]_{(3)} = \text{diag}(\ma{g}_i) \in \bb{C}^{M_TN \times M_TN}$. Defining 
\begin{equation}
    \ma{\Omega} \doteq \Big(\ma{X} \otimes \ma{HS}\Big) \in \bb{C}^{TM_R \times M_TN},
    \label{EQ: Omega definition}
\end{equation}
and 
\begin{equation}
    \ma{\Psi}^{\text{T}} \doteq \Big(\ma{W}^{\text{T}} \diamond \ma{P}^{\text{T}}\Big) \in \bb{C}^{M_TN \times K},
    \label{EQ: PSI definition}
\end{equation}
we have
\begin{equation}
 [\ten{Y}_i]_{(3)}^\text{T} = \ma{\Omega}[\ten{T}_{\ma{G}_i}]_{(3)}\ma{\Psi}^{\text{T}} \in \bb{C}^{TM_R \times K}\;.
 \label{EQ:frontal_sice PARAFAC vision}
\end{equation}
Let $\ten{Z} \in \bb{C}^{TM_R \times K \times I}$ be a third order tensor which is defined as
\begin{equation}
    \ten{Z} \doteq [\ten{Y}_1]_{(3)}^\text{T} \sqcup_3 [\ten{Y}_2]_{(3)}^\text{T} \dots \sqcup_3 [\ten{Y}_I]_{(3)}^\text{T},
\end{equation}
such that
\begin{equation}
    \ten{Z} = \ten{I}_{3,M_TN} \times_1 \ma{\Omega} \times_2 \ma{\Psi} \times_3 \overline{\ma{G}},
    \label{EQ: produto modo N de Z}
\end{equation}
where $\overline{\ma{G}} = \Big[\ma{g}_1 \; \dots \; \ma{g}_I\Big]^{\text{T}} \in \bb{C}^{I \times M_TN}$. Note that $\ten{Z}$ matches a third-order PARAFAC tensor model \cite{Harshman}. In Figure \ref{fig:PARAFAC_Block}, we illustrate the procedures that have been performed until now. It can be seen that the received signal model $\ma{Y}_{i,k}$ is a PARATUCK(2,4) tensor, denoted by $\ten{Y}$. Concatenating this tensor through the vectorized version related to the $3$-mode $K$,  for each frame we have a third-order tensor $\ten{Y}_i \in \bb{C}^{M_R \times T \times K}$ that matches a third-order TUCKER model, $i = 1,\ldots, I$. However, the $3$-mode unfolding transposed of $\ten{Y}_i$ can be seen as the $i$-th frontal slice of a tensor $\ten{Z} \in \bb{C}^{M_RT \times K \times I}$ that fits a PARAFAC model where the factor matrices are structured. Recall that $\ma{\Omega}$ is the  Kronecker product between the data symbols $\ma{X}$ and the equivalent channel $\ma{HS}$, while $\ma{\Psi}$ is the Khatri-Rao product (column-wise Kronecker) between the rotation matrix $\ma{P}$ and the coding matrix $\ma{W}$. In this sense, the following correspondence can be established:

\begin{figure}[!t]
    \centering
    \includegraphics[scale = 0.49]{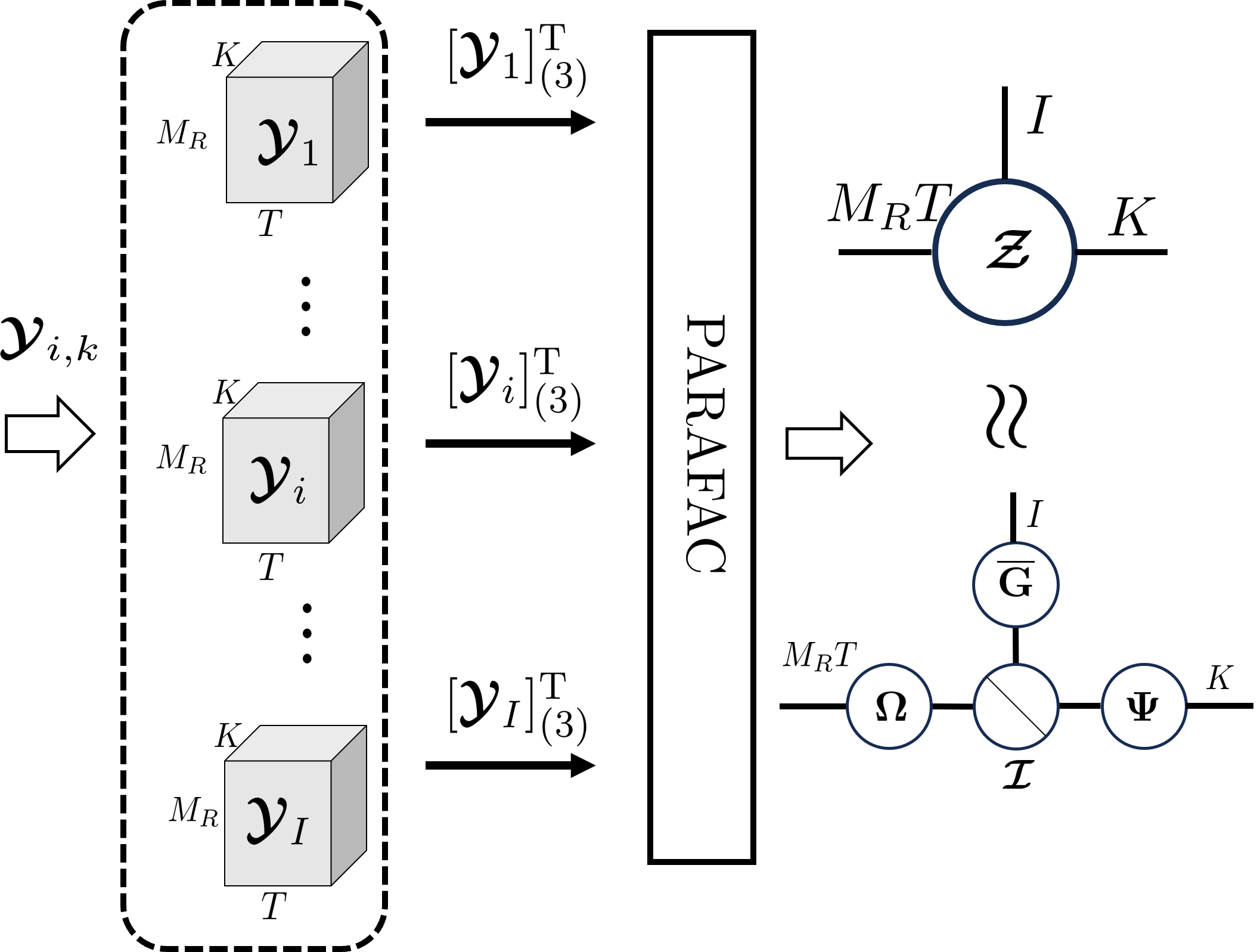}
    \caption{Third-order PARAFAC tensor model: The received signal at $(14)$ fits a fourth-order PARATUCK tensor model. From the third mode unfolding of the tensor $\ten{Y}_{i,k}$, we derive a third-order PARAFAC model, with $\ma{\Omega}$, $\ma{\Psi}$ and $\overline{\ma{G}}$ as factor matrices.}
    \label{fig:PARAFAC_Block}
\end{figure}
\begin{equation}
    \begin{aligned}
        \Big(\ma{A}, \; \ma{B}, \; \ma{C} \Big) & \leftrightarrow \Big(\ma{\Omega}, \; \ma{\Psi}, \; \overline{\ma{G}} \Big)\\
        \Big(I_1, \; I_2, \; I_3 \Big) & \leftrightarrow \Big(TM_R, \; K, \; I \Big).
    \end{aligned}
\end{equation}
To estimate the factor matrix $\ma{\Omega}$  and  the channel $\overline{\ma{G}}$, we must solve the following optimization problem
\begin{equation}
\resizebox{0.48\textwidth}{!}{
$\Big(\hat{\overline{\ma{G}}}, \; \hat{\ma{\Omega}}\Big)
= \underset{\overline{\ma{G}}, \ma{\Omega}}{\arg\min}
\left\|\ten{Z} - \ten{I}_{3,M_TN}
\times_1 \ma{\Omega}
\times_2 \ma{\Psi}
\times_3 \overline{\ma{G}}\right\|_F^2$.}
\label{Eq:nonlinear problem}
\end{equation}

Note that (\ref{Eq:nonlinear problem}) is a nonlinear problem due to the presence of two unknown matrices $(\overline{\ma{G}} \; \text{and} \; \ma{\Omega})$. To address this issue, we can derive sub-optimization linear programs in which certain matrices are held constant. For example, by fixing the channel matrix $\overline{\ma{G}}$, the problem becomes linear in the factor matrix $\ma{\Omega}$. Similarly, fixing $\ma{\Omega}$ we have a linear problem in $\overline{\ma{G}}$. Consequently, the optimization problem in (\ref{Eq:nonlinear problem}) can be decomposed into sub-optimization problems, which can then be solved iteratively. This is the underlying premise of the \ac{ALS} algorithm.  On the other hand, since we are interested in solving only for $\ma{\Omega}$ and $\overline{\ma{G}}$, the algorithm is defined as \ac{BALS}. For this, we make use of $[\ten{Z}]_{(1)}$ and $[\ten{Z}]_{(3)}$, which, from (\ref{EQ:frontal_sice PARAFAC vision}), are defined as  
\begin{equation}
    \begin{aligned}
    \relax[\ten{Z}]_{(1)} &=  \Big[  [\ten{Y}_1]_{(3)}^\text{T} \;  [\ten{Y}_2]_{(3)}^\text{T} \; \dots \;  [\ten{Y}_I]_{(3)}^\text{T} \Big]\\
    & = \ma{\Omega}\Big[\text{diag}(\ma{g}_1)\ma{\Psi}^{\text{T}} \; \text{diag}(\ma{g}_2)\ma{\Psi}^{\text{T}} \; \dots \; \text{diag}(\ma{g}_I)\ma{\Psi}^{\text{T}}\Big]\\
    & = \ma{\Omega}\Big( \overline{\ma{G}} \diamond \ma{\Psi} \Big)^{\text{T}}\; \in \bb{C}^{TM_R \times IK}\; ,
    \label{EQ: 1-mode unfoldingTilde_Y as PARAFAC}
    \end{aligned}
\end{equation}
and
\begin{equation}
    \begin{aligned}
    \relax[\ten{Z}]_{(3)} &=  \Big[ \text{vec}( [\ten{Y}_1]_{(3)}^\text{T}) \; \text{vec}( [\ten{Y}_2]_{(3)}^\text{T}) \; \dots \; \text{vec}( [\ten{Y}_I]_{(3)}^\text{T}) \Big]^{\text{T}}\\
    & = \Big[\big(\ma{\Psi} \diamond \ma{\Omega}\big)[\ma{g}_1 \; \ma{g}_2 \; \dots \; \ma{g}_I]\Big]^\text{T}\\
    & = \overline{\ma{G}}\big(\ma{\Psi}  \diamond \ma{\Omega}\big)^{\text{T}} \in \bb{C}^{I \times KTM_R}.
    \label{EQ: 3-mode unfoldingTilde_Y as PARAFAC}
    \end{aligned}
\end{equation}
Using the first and third modes unfoldings, as described in equations (\ref{EQ: 1-mode unfoldingTilde_Y as PARAFAC}) and (\ref{EQ: 3-mode unfoldingTilde_Y as PARAFAC}), respectively, we propose an iterative solution to estimate the structured matrix $\ma{\Omega}$ and the channel matrix $\overline{\ma{G}}$. It is important to note that after estimating $\ma{\Omega}$, a Kronecker factorization problem must be solved to recover the data symbol matrix $\ma{X}$ and the BS-IRS channel $\ma{H}$. This yields a two-stage solution: first, we solve the PARAFAC problems using the \ac{BALS} algorithm, followed by a Kronecker factorization. Given this combined approach, we have named this algorithm PAKRON (Algorithm \ref{Algo:PAKRON}).
\begin{algorithm}[!t]
\small
\IncMargin{1.5em} 
\DontPrintSemicolon
\SetKwInOut{Input}{Input}\SetKwInOut{Output}{Output}

\Input{Received signal $\mathcal{Y}$}
\Output{$\hat{\mathbf{X}}$, $\hat{\mathbf{H}}$, and $\hat{\overline{\mathbf{G}}}$}

\BlankLine
\textbf{Procedure:}\;
\vspace{0.3em}

\text{1. Execute Stage I using Algorithm \ref{Algo: PseudocodeBALS_PARAFAC}:}\;
\textit{Obtain initial estimates $\hat{\mathbf{\Omega}}$ and $\hat{\overline{\mathbf{G}}}$.}\;
\vspace{0.5em}

\text{2. Execute Stage II using Algorithm \ref{Algo:Kron}:}\;
\quad  \textit{Decouple $\hat{\mathbf{\Omega}}$ to obtain final estimates $\hat{\mathbf{X}}$ and $\hat{\mathbf{H}}$.}\;

\vspace{0.5em}
\caption{PAKRON receiver summary}
\label{Algo:PAKRON}
\end{algorithm}
\vspace{-0.5cm}
\subsubsection{PAKRON receiver (Stage I)}
\label{sec: PACKRON receiver}
 Capitalizing on the matrix unfoldings (\ref{EQ: 1-mode unfoldingTilde_Y as PARAFAC}) and (\ref{EQ: 3-mode unfoldingTilde_Y as PARAFAC}), we derive the \ac{BALS} algorithm. This algorithm estimates the matrices $\ma{\Omega}$ and $\overline{\ma{G}}$ in an alternating manner by iteratively optimizing the following two cost functions:
\begin{eqnarray}
&\hat{\ma{\Omega}} = \underset{\ma{\Omega}}{\arg\min} \,\, \left\|[\ten{Y}]_{(1)} - \ma{\Omega}\Big( \overline{\ma{G}} \diamond \ma{\Psi}\Big)^{\textrm{T}}\right\|_F^2\label{func costG},\\
&\hat{\overline{\ma{G}}}^\text{T} = \underset{\overline{\ma{G}}}{\arg\min}\,\, \left\|[\ten{Y}]_{(3)} - \overline{\ma{G}}^\text{T}\Big(\ma{\Psi} \diamond \ma{\Omega}\Big)^{\textrm{T}}\right\|_F^2,
\label{Func costZ}
\end{eqnarray}
the solutions of which are, respectively, given by
\begin{eqnarray}
&\hat{\ma{\Omega}} =[\ten{Y}]_{(1)}\Big[\Big( \overline{\ma{G}} \diamond \ma{\Psi}\Big)^{\textrm{T}} \Big]^\dagger,\label{EstimaG}\\
&\hat{\overline{\ma{G}}} =  [\ten{Y}]_{(3)}\Big[\Big(\ma{\Psi} \diamond \ma{\Omega} \Big)^{\textrm{T}}\Big]^\dagger. \label{EstimaH}
\end{eqnarray}
\begin{algorithm}[!t]
\small
\IncMargin{0.1em}
\DontPrintSemicolon
\SetKwInOut{Input}{Input}\SetKwInOut{Output}{Output}

\Input{Received signal $\mathcal{Y}$} 
\Output{$\hat{\mathbf{\Omega}}$, $\hat{\overline{\mathbf{G}}}$}

\BlankLine
Initialize $i = 0$ and $\hat{\overline{\mathbf{G}}}_0$ randomly\;

\Repeat{$\|e(i) - e(i-1)\| < \delta$}{
    $i = i + 1$\;
    
    \text{1. Estimate $\hat{\mathbf{\Omega}}_i$ via Least Squares:}\;
    $\quad \hat{\mathbf{\Omega}}_i = [\mathcal{Y}]_{(1)} [(\hat{\overline{\mathbf{G}}}_{i-1} \diamond \mathbf{\Psi})^{\text{T}}]^\dagger$\;
    \vspace{0.3em}
    
    \text{2. Estimate $\hat{\overline{\mathbf{G}}}_i$ via Least Squares:}\;
    $\quad \hat{\overline{\mathbf{G}}}_i = [\mathcal{Y}]_{(3)} [(\mathbf{\Psi} \diamond \hat{\mathbf{\Omega}}_i)^{\text{T}}]^\dagger$\;
    \vspace{0.3em}
}

\caption{Stage I of the PAKRON receiver}
\label{Algo: PseudocodeBALS_PARAFAC}
\end{algorithm}
The convergence of Algorithm \ref{Algo: PseudocodeBALS_PARAFAC} is declared when $\|e_{(i)} - e_{(i-1)}\|\leq \delta$, where $e_{(i)} = \|\mathcal{Y} - \hat{\mathcal{Y}}_{(i)}\|_{F}^{2}$ denotes the reconstruction error computed at the $i$-th iteration. $\delta$ is a predefined threshold parameter, and $\hat{\mathcal{Y}}_{(i)}=[\hat{\overline{\ma{G}}}_{(i)}, \hat{\ma{\Omega}}_{(i)}, \ma{\Psi}]$ is the reconstructed PARAFAC model at the $i$-th iteration. From Stage I, we obtain an estimate of $\ma{\Omega}$, as defined in equation (\ref{EQ: Omega definition}), which is subsequently used in Stage II.
\vspace{-0.5cm}
\subsubsection{PAKRON receiver (Stage II)}

Once the estimate of $\bm{\Omega}$ is found in Stage I, the estimation of the data symbol matrix $\ma{X}$ and the RIS-BS channel $\bm{H}$ is achieved in stage II by solving the following problem:
\begin{equation}
    (\hat{\ma{X}}, \hat{\ma{H}}) = \underset{\ma{X}, \ma{H}}{\arg\min}\|\hat{\ma{\Omega}} - \ma{X} \otimes (\ma{HS})\|_{\text{F}}^{2}.
    \label{EQ:Cost_Func_Kron_Omega}
\end{equation}
Considering the unitary design of the BD-RIS design and the properties (\ref{prop: Kron_product}) and (\ref{prop: Kron_Inverse}), we get $\ma{\Delta} = \hat{\ma{\Omega}}(\ma{I}_{M_T} \otimes \ma{S}^{\text{H}})$. Then, the problem in $(\ref{EQ:Cost_Func_Kron_Omega})$ is equivalent to 
\begin{equation}
    (\hat{\ma{X}}, \hat{\ma{H}}) = \underset{\ma{X}, \ma{H}}{\arg\min}\|\ma{\Delta} - \ma{X} \otimes \ma{H}\|_{\text{F}}^{2}.
    \label{EQ:Cost_Func_Kron_Delt}
\end{equation}
The matrix $\ma{\Delta}$ can be properly rearranged, in which case the problem (\ref{EQ:Cost_Func_Kron_Delt}) can be solved by means of rank-one matrix approximation problems, as follows
\begin{equation}
    (\hat{\ma{X}}, \hat{\ma{H}}) = \underset{\ma{X}, \ma{H}}{\min}\|\widetilde{\ma{Z}} - \ma{x}\ma{h}^{\textrm{T}}\|_\text{F},
    \label{EQ:Cost Func KrD _Rank_One}
\end{equation}
where 
\begin{equation}
	\begin{split}
 	\tilde{\ma{Z}} & = \left[\begin{array}{ccc}
 		\ma{\Delta}_{1,1} \, &
 		\dots \,
 		&
 		\ma{\Delta}_{T,Mr}
 	\end{array}\right]^{\text{T}} \in \bb{C}^{TM_T \times M_RN}\; ,
 \label{EQ: Q_Tilde}
   \end{split}
\end{equation}
and $\ma{\Delta}_{t,m_t}  = x_{t,m}\ma{H} \in \bb{C}^{M_r \times N}$ with $t = 1, \dots, T$ and $m_t = 1,  \dots, M_t$. $\ma{x} = \text{vec}(\ma{X})$ and $\ma{h} = \text{vec}(\ma{H})$. 
According to \cite{Eckart_36}, the optimal solution for the \ac{RIS}-BS channel and the data symbol matrix is obtained by taking the left and right singular vectors of $\tilde{\ma{Z}}$, as illustrated in Algorithm \ref{Algo:Kron}.\footnote{This ambiguity-removal strategy is consistent with semi-blind tensor-based receivers that rely on a small number of known symbols to calibrate scaling factors \cite{Du2023SemiBlindRIS}.}
\begin{algorithm}[!t]
\IncMargin{1em}
	\DontPrintSemicolon
	\DontPrintSemicolon
	\SetKwData{Left}{left}\SetKwData{This}{this}\SetKwData{Up}{up}
	\SetKwFunction{Union}{Union}\SetKwFunction{FindCompress}{FindCompress}
	\SetKwInOut{Input}{input}\SetKwInOut{Output}{output}
	\textbf{Procedure}\newline
	\Input{$\hat{\ma{\Omega}}$}
	\Output{$\hat{\ma{X}}$ and $\hat{\ma{H}}$ }
	\BlankLine
	\Begin{
	\begin{enumerate}
	    \item[1.] \textit{Construct the rank-one matrix} $\tilde{\ma{Z}} \in \bb{C}^{TM_T \times M_RN}$
	    \item[2.] $(\ma{u}_1,\ma{\sigma}_1,\ma{v}_1)\longleftarrow\textrm{truncated-SVD}(\tilde{\ma{Z}})$\;
	    $\hat{\ma{x}} \longleftarrow \sqrt{\sigma_1}\ma{u}_1$;\quad
	    $\hat{\ma{h}} \longleftarrow \sqrt{\sigma_1}\ma{v}_{1}^{\ast}$
	    \item[3.] \textit{Reconstruct} $\hat{\ma{X}}$ \text{and} $\hat{\ma{H}}$ \textit{by unvec} $\hat{\ma{x}}$ \textit{and} $\hat{\ma{h}}$
	    \item[4.] \textit{Remove the scaling ambiguities of} $\hat{\ma{X}}$ \textit{and} $\hat{\ma{H}}$.
	\end{enumerate}
	\textbf{end}
	    }
    \caption{Stage II of the PAKRON receiver}
	\label{Algo:Kron}
\end{algorithm}

\subsection{TUCKER-Based Receiver}
\begin{figure}[!t]
    \centering
    \includegraphics[scale=0.5]{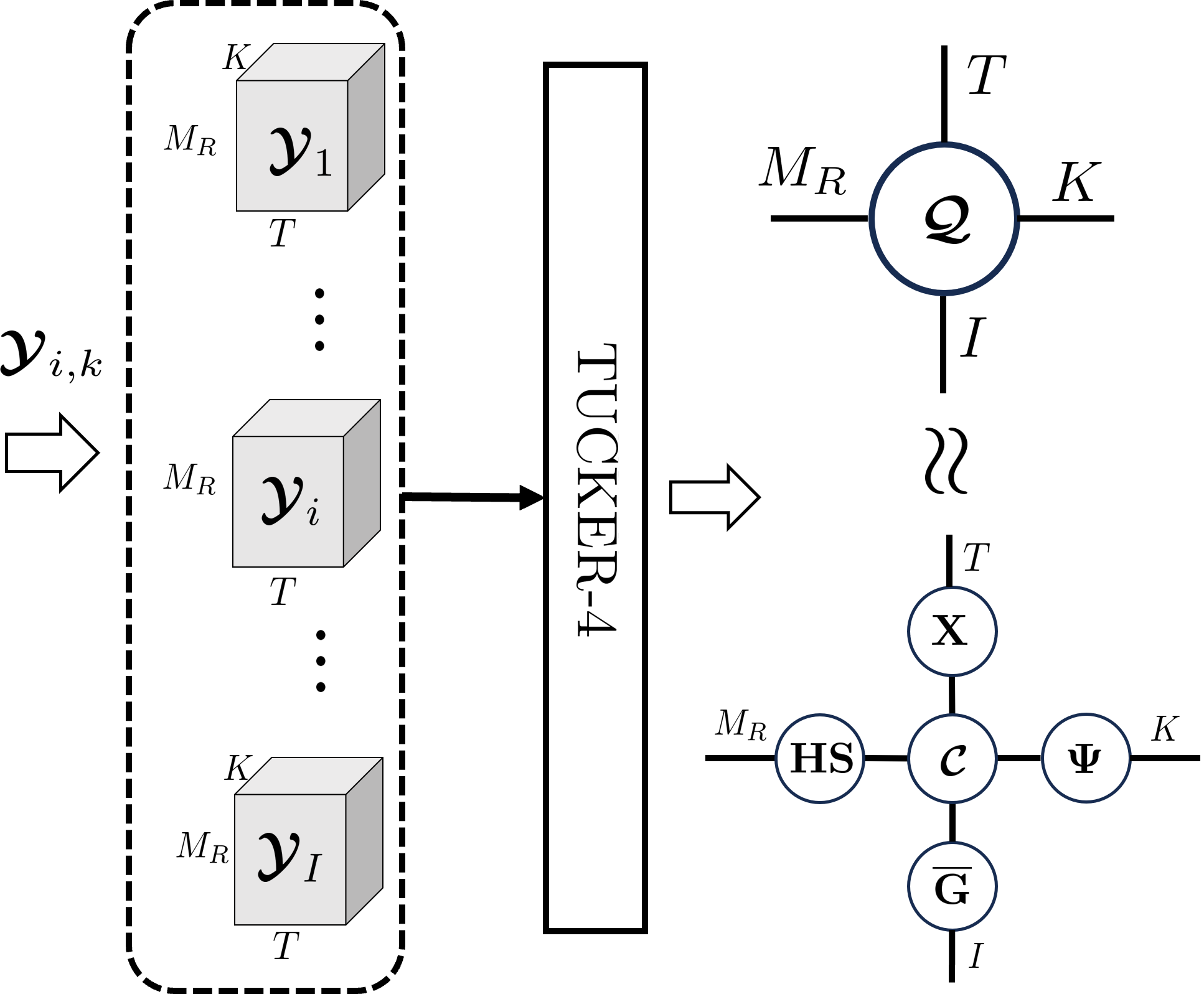}
    \caption{TUCKER tensor model: In this scenario, the tensor from the $i$-th frame, $\ten{Y}_i$ in $(18)$, is rearranged to derive a fourth-order TUCKER model for the received signal tensor.}
    \label{fig:PT24 as TUCKER4}
\end{figure}

The approach developed in Section \ref{Sec: PARAFAC design} combines the temporal domain (number of time slots, $T$) with the spatial domain (number of receiver antennas, $M_R$), yielding a structured PARAFAC model. As a consequence, a two-stage joint symbol and channel estimation is performed, which can lead to error propagation and degrade system performance in noise-limited scenarios. However, alternative strategies can be applied to rearrange the received signal, as illustrated in Figure \ref{fig:PT24 as TUCKER4}, where two temporal dimensions (frames $I$ and blocks $K$) are combined.
Then, let us revisit the received signal at (\ref{EQ: Y_i})
\begin{equation}
\ma{Y}_{i}  = \Big(\ma{X} \otimes \ma{HS}\Big)\text{diag}(\ma{g}_i)\ma{\Psi}^{\text{T}}\;,
\label{EQ: Y_i_tucker vision}
\end{equation}
and its $1$-mode unfolding in (\ref{EQ: 1-mode unfoldingTilde_Y as PARAFAC})
\begin{equation}
    \begin{aligned}
    \relax[\ten{Y}]_{(1)} &=  \Big[ \ma{Y}_1 \; \ma{Y}_2 \; \dots \; \ma{Y}_I \Big]\\
    & = \Big(\ma{X} \otimes \ma{HS}\Big)\Big( \overline{\ma{G}} \diamond \ma{\Psi} \Big)^{\text{T}}\; \in \bb{C}^{TM_R \times IK}\; .
    \label{EQ: 1-mode unfoldingTilde_Y as PARAFAC/Tucker}
    \end{aligned}
\end{equation}
Note that equation (\ref{EQ: 1-mode unfoldingTilde_Y as PARAFAC/Tucker}) can be seen as a third order tensor $\widetilde{\ten{Y}} \in \bb{C}^{M_R \times T \times IK}$ that corresponds to a TUCKER model where the $n$-mode product is
\begin{equation}
    \widetilde{\ten{Y}} = (\overline{\ten{G}} \diamond_3 \ten{T}) \times_1 \ma{HS} \times_2 \ma{X} \times_3 \ma{I}_{IK}\; \in \bb{C}^{M_R \times T \times IK}\; ,
    \label{EQ: N-mode notation for tucker notation}
\end{equation}
where the core tensor $\ten{C} = \Big(\overline{\ten{G}} \diamond_3 \ten{T}\Big) \in \bb{C}^{N \times M_T \times IK}$ has a special structure which is given by $3$-mode Khatri-Rao product between two tensors, which are 
\begin{equation}
    \overline{\ten{G}} \doteq \Big[\overline{\ma{G}}_1 \sqcup_3 \; \overline{\ma{G}}_2 \sqcup_3 \; \dots \; \sqcup_3 \overline{\ma{G}_I} \Big] \; \bb{C}^{N \times M_T \times I}
    \label{EQ:Core tensor Gbar}
\end{equation}
and
\begin{equation}
    \ten{T}  \doteq \ten{I}_{3,K} \times_1 \ma{P}^{\textrm{T}} \times_2 \ma{W}^{\textrm{T}} \times_3 \ma{I}_K \in \bb{C}^{N \times M_T \times K}\;,
    \label{EQ:core tensor T}
\end{equation}
where $\sqcup_3$ indicates concatenation along the third dimension. Note that the channel $\overline{\ma{G}}$ lies within the core tensor. This implies that, in a straightforward approach, one would need at least a two-stage procedure: first, estimate the core tensor and then extract $\overline{\ma{G}}$. 
Instead of reducing the tensor order, however, we can preserve the fourth-order structure and reinterpret it as a TUCKER model. This can be done by applying the multimodal unfolding concept, which shows that the PARATUCK-(2,4) model can be expressed as a fourth-order TUCKER tensor. This relationship is depicted in Fig. \ref{fig:PT24 as TUCKER4}, where the $I$ tensors $\ten{Y}_i$ ($i=1,\ldots,I$) are transformed, by multimode unfolding, into a TUCKER tensor $\ten{Q}$ (in tensor network notation). This transformation preserves the original dimensions, but changes the underlying tensor model. Such a reformulation enables the derivation of a single-stage solution, as discussed next.
 
We know that given $\ma{A} \in \bb{C}^{N_A \times L}$ and $\ma{B} \in \bb{C}^{N_B \times L}$, the Khatri-Rao product between them can be expressed as $\ma{A} \diamond \ma{B} = (\ma{A} \otimes \ma{B})\ma{\Xi} \in \bb{C}^{N_AN_B \times L}$ where $\ma{\Xi}$ is an $L^2 \times L$ selection matrix that extracts the column-wise Kronecker product. Now, applying this Khatri-Rao and Kronecker product relationship to the $1$-mode unfolding in (\ref{EQ: 1-mode unfoldingTilde_Y as PARAFAC/Tucker}), we can rewrite it as
\begin{equation}
    \begin{aligned}
    \relax[\ten{Y}]_{(1)} & = \Big(\ma{X} \otimes \ma{HS}\Big)\Big( \overline{\ma{G}} \diamond \ma{\Psi} \Big)^{\text{T}}\; \in \bb{C}^{TM_R \times IK}\;\\
    & = \Big(\ma{X} \otimes \ma{HS}\Big)\Big((\overline{\ma{G}} \otimes \ma{\Psi}) \ma{\Xi}\Big)^{\text{T}}\\
    & = \Big(\ma{X} \otimes \ma{HS}\Big)\ma{\Xi}^{\text{T}}\Big(\overline{\ma{G}} \otimes \ma{\Psi}\Big)^{\text{T}},
    \label{EQ: 1-mode unfoldingTilde_Y as Tucker}
    \end{aligned}
\end{equation}
where $\ma{\Xi} \in \bb{C}^{(M_TN)^2 \times M_TN}$ is the selection matrix such that $\ma{\Xi} = \ma{I}_{M_TN} \diamond \ma{I}_{M_TN}$. From these observations, we can note that $ [\ten{Y}]_{(1)}$ can be seen as a generalized unfolding of a tensor $\ten{Q}$, that is, $[\ten{Q}]_{([1,2],[3,4])} = [\ten{Y}]_{(1)}$. Then
\begin{equation}
    \begin{aligned}
    \relax[\ten{Q}]_{([1,2],[3,4])}  & = \Big(\ma{X} \otimes \ma{HS}\Big)(\ma{I}_{M_TN} \diamond \ma{I}_{M_TN})^{\text{T}}\Big(\overline{\ma{G}} \otimes \ma{\Psi}\Big)^{\text{T}},
    \label{EQ: multi-mode unfolding Y Tucker}
    \end{aligned}
\end{equation}
where $(\ma{I}_{M_TN} \diamond \ma{I}_{M_TN})^{\text{T}} = [\ten{C}]_{([1,2],[3,4])} \in \bb{C}^{M_TN \times (M_TN)^2}$ is the multimode unfolding of a fourth order core tensor $\ten{C} \in \bb{C}^{N \times M_T \times M_TN \times M_TN}$, while equation (\ref{EQ: multi-mode unfolding Y Tucker}) denotes the multimode unfolding of the tensor $\ten{Q} \in \bb{C}^{M_R \times T \times K \times I}$, whose $n$-mode product notation is
\begin{equation}
    \ten{Q} = \ten{C} \times_1 \ma{HS} \times_2 \ma{X} \times_3 \ma{\Psi} \times_4 \overline{\ma{G}}.
    \label{EQ: product mode n of tucker model}
\end{equation}
Comparing (\ref{EQ: product mode n of tucker model}) with (\ref{eq:tenZ_tucker_generic}), the fourth-order tensor $\ten{Q}$ fits a TUCKER model, where the matrix $\ma{\Psi}$ is known. To estimate the matrices $\ma{HS}$, $\ma{X}$ and $\overline{\ma{G}}$ it is necessary to solve the following optimization problem:
\begin{equation}
    \underset{(\ma{HS}, \ma{X}, \overline{\ma{G}})}{\arg\min}\| \ten{Q} - \ten{C} \times_1 \ma{HS} \times_2 \ma{X} \times_3 \ma{\Psi} \times_4 \overline{\ma{G}}\|_F^2.
    \label{EQ: General opt problem}
\end{equation}
The problem in (\ref{EQ: General opt problem}) is nonlinear since the three factor matrices are unknown. However,
it can be unfolded into a fourth-order model formulation in terms of its factor matrices as follows:
\begin{equation}
    [\ten{Q}]_{(1)} = \ma{HS}[\ten{C}]_{(1)}\big(\overline{\ma{G}} \otimes \ma{\Psi} \otimes \ma{X}\big)^{\text{T}} \in \bb{C}^{M_R \times IKT},
    \label{EQ:1-mode tucker}
\end{equation}
\begin{equation}
    [\ten{Q}]_{(2)} = \ma{X}[\ten{C}]_{(2)}\big(\overline{\ma{G}} \otimes \ma{\Psi} \otimes \ma{HS}\big)^{\text{T}} \in \bb{C}^{T \times IKM_R},
    \label{EQ:2-mode tucker}
\end{equation}
\begin{equation}
    [\ten{Q}]_{(3)} = \ma{\Psi}[\ten{C}]_{(3)}\big(\overline{\ma{G}} \otimes \ma{X} \otimes \ma{HS}\big)^{\text{T}} \in \bb{C}^{K \times ITM_R},
    \label{EQ:3-mode tucker}
\end{equation}
\begin{equation}
    [\ten{Q}]_{(4)} =\overline{\ma{G}}[\ten{C}]_{(4)}\big(\ma{\Psi} \otimes \ma{X} \otimes \ma{HS}\big)^{\text{T}} \in \bb{C}^{I \times KTM_R},
    \label{EQ:4-mode tucker}
\end{equation}
with $[\ten{C}]_{(1)} \in \bb{C}^{N \times (M_TN)^2M_T}$, $[\ten{C}]_{(2)} \in \bb{C}^{M_T \times (M_TN)^2N}$, $[\ten{C}]_{(3)} \in \bb{C}^{M_TN \times (M_TN)^2}$ and $[\ten{C}]_{(4)} \in \bb{C}^{M_TN \times (M_TN)^2}$.
From the first-, second-, and fourth mode unfoldings, we can derive three sub-optimization linear problems as given 
\begin{equation}
    \underset{(\ma{HS})}{\arg\min}\| [\ten{Q}]_{(1)} - \ma{HS}[\ten{C}]_{(1)}\big(\overline{\ma{G}} \otimes \ma{\Psi} \otimes \ma{X}\big)^{\text{T}}\|_F^2,
    \label{EQ: OPT problem to HS}
\end{equation}
\begin{equation}
    \underset{(\ma{X})}{\arg\min}\|  [\ten{Q}]_{(2)} - \ma{X}[\ten{C}]_{(2)}\big(\overline{\ma{G}} \otimes \ma{\Psi} \otimes \ma{HS}\big)^{\text{T}} \|_F^2,
    \label{EQ: OPT problem to X}
\end{equation}
\begin{equation}
    \underset{(\overline{\ma{G}})}{\arg\min}\|   [\ten{Q}]_{(4)} - \overline{\ma{G}}[\ten{C}]_{(4)}\big(\ma{\Psi} \otimes \ma{X} \otimes \ma{HS}\big)^{\text{T}} \|_F^2.
    \label{EQ: OPT problem to Gbar}
\end{equation}
To estimate $\ma{HS}$, $\ma{X}$ and $\overline{\ma{G}}$, we alternately solve the \ac{LS} problems in (\ref{EQ: OPT problem to HS}), (\ref{EQ: OPT problem to X}) and (\ref{EQ: OPT problem to Gbar}), respectively, whose solutions are given by:
\begin{equation}
    \hat{\ma{HS}} = [\ten{Q}]_{(1)}\left[[\ten{C}]_{(1)}\big(\overline{\ma{G}} \otimes \ma{\Psi} \otimes \ma{X}\big)^{\text{T}}\right]^{\dagger}
    \label{EQ: Estimate HS}
\end{equation}
\begin{equation}
    \hat{\ma{X}} = [\ten{Q}]_{(2)}\left[[\ten{C}]_{(2)}\big(\overline{\ma{G}} \otimes \ma{\Psi} \otimes \ma{HS}\big)^{\text{T}}\right]^{\dagger}
    \label{EQ: Estimate X}
\end{equation}
\begin{equation}
    \hat{\overline{\ma{G}}} = [\ten{Q}]_{(4)}\left[[\ten{C}]_{(4)}\big(\ma{\Psi} \otimes \ma{X} \otimes \ma{HS}\big)^{\text{T}}\right]^{\dagger}.
    \label{EQ: Estimate Gbar}
\end{equation}
This procedure is the foundation of the trilinear alternating least squares TUCKER receiver described in the Algorithm \ref{Algo: Pseudocode TALS-TUCKER}. The convergence criterion is similar to the Algorithm \ref{Algo: PseudocodeBALS_PARAFAC}. Note that this approach avoids propagating error from the two-stage PAKRON solution.
\begin{algorithm}[!t]
\small
\DontPrintSemicolon
\SetKwInOut{Input}{Input}
\SetKwInOut{Output}{Output}

\Input{Received signal tensor $\mathcal{Q}$}
\Output{$\hat{\mathbf{H}}$, $\hat{\overline{\mathbf{G}}}$, and $\hat{\mathbf{X}}$}

\BlankLine
Initialize $i = 0$, $\hat{\overline{\mathbf{G}}}_{0}$ and $\hat{\mathbf{X}}_{0}$ randomly\;

\Repeat{$\|e(i) - e(i-1)\| < \delta$}{
    $i = i + 1$\;
    
     \textit{1. Estimate $\hat{\mathbf{HS}}_{i}$ via Least Squares:}\;
    \Indp $\hat{(\mathbf{HS})}_i = [\mathcal{Q}]_{(1)} \left[ [\mathcal{C}]_{(1)} (\hat{\overline{\mathbf{G}}}_{i-1} \otimes \mathbf{\Psi} \otimes \hat{\mathbf{X}}_{i-1})^{\text{T}} \right]^{\dagger}$\;
    \Indm \vspace{0.3em}

     \textit{2. Estimate $\hat{\mathbf{X}}_i$ via Least Squares:}\;
    \Indp $\hat{\mathbf{X}}_i = [\mathcal{Q}]_{(2)} \left[ [\mathcal{C}]_{(2)} (\hat{\overline{\mathbf{G}}}_{i-1} \otimes \mathbf{\Psi} \otimes \hat{(\mathbf{HS})}_i)^{\text{T}} \right]^{\dagger}$\;
    \Indm \vspace{0.3em}

     \textit{3. Estimate $\hat{\overline{\mathbf{G}}}_i$ via Least Squares:}\;
    \Indp $\hat{\overline{\mathbf{G}}}_i = [\mathcal{Q}]_{(4)} \left[ [\mathcal{C}]_{(4)} (\mathbf{\Psi} \otimes \hat{\mathbf{X}}_i \otimes \hat{(\mathbf{HS})}_i)^{\text{T}} \right]^{\dagger}$\;
    \Indm \vspace{0.3em}
}

\caption{TUCKER receiver (T-ALS)}
\label{Algo: Pseudocode TALS-TUCKER}
\end{algorithm}
\textit{Remark 2:} From (\ref{EQ:3-mode tucker}), an alternative (closed-form) approach can be used to jointly estimate the involved channels $\ma{H}$, $\ma{G}$, as well as the data symbol matrix $\ma{X}$. This approach left-filters the $3$-mode unfolding of the received signal using $(\ma{\Psi}[\ten{C}]_{(1)})^{\dagger}$ followed by solving a ``triple-Kronecker'' factorization problem of the filtered signal. The algorithm is similar to the one proposed in \cite{SOKAL_Elsevier}. Although this approach is simpler, it imposes a stricter constraint on the length of the data block, requiring $K \geq (M_TN)^2$.

\section{IDENTIFIABILITY AND COMPLEXITY}
\label{Sec: Indentifiability}
In this section, we discuss the identifiability conditions for our proposed solutions. The study provides a useful relationship among the system parameters necessary to obtain unique estimates of the channel matrices and the symbol matrix when using the proposed semi-blind receivers. 

\subsection{TUCKER receiver} For the TUCKER receiver, the problems (\ref{EQ: OPT problem to HS}), (\ref{EQ: OPT problem to X}), and (\ref{EQ: OPT problem to Gbar}) are unique in the \ac{LS} sense if $\ma{V}_1 = [\ten{C}]_{(1)}\big(\overline{\ma{G}} \otimes \ma{\Psi} \otimes \ma{X}\big)^{\text{T}} \in \bb{C}^{N \times IKT}$, $\ma{V}_2 = [\ten{C}]_{(2)}\big(\overline{\ma{G}} \otimes \ma{\Psi} \otimes \ma{HS}\big)^{\text{T}} \in \bb{C}^{M_T \times IKM_R}$ and $\ma{V}_4 = [\ten{C}]_{(4)}\big(\ma{\Psi} \otimes \ma{X} \otimes \ma{HS}\big)^{\text{T}} \in \bb{C}^{M_TN \times KTM_R}$ are full row rank. Satisfying these conditions   implies the following inequalities
\begin{equation}
    IKT \geq N\; ,\ \; IKM_R \geq M_T \;  \text{and}\; KTM_R \geq M_TN.
\end{equation}
Combining them, we obtain a necessary condition for the uniqueness of the TUCKER receiver in the \ac{LS} sense, in terms of the required number of blocks as
\begin{equation}
    K \geq \max\Big(\dfrac{N}{IT} , \dfrac{M_T}{IM_R}, \dfrac{M_TN}{TM_R}\Big).
    \label{EQ: condition to uniqueness for TALS-TUCKER}
\end{equation}

\subsubsection{PAKRON receiver}
The conditions necessary to ensure a unique joint channel and symbol recovery for the PAKRON receiver are associated with stage I. In this sense,  the problems (\ref{func costG}) and (\ref{Func costZ}) are unique in the LS sense if $\ma{R}_1 = \overline{\ma{G}} \diamond \ma{\Psi} \in \bb{C}^{IK \times M_TN}$ and $\ma{R}_2 = \ma{\Psi} \diamond \ma{\Omega} \in \bb{C}^{KTM_R \times M_TN}$ are full column rank, which implies the following conditions to be satisfied:

\begin{equation}
    IK \geq M_TN \quad \text{and} \quad KTM_R \geq M_TN,
\end{equation}
or, equivalently,
\begin{equation}
  K \geq \dfrac{M_TN}{I} \quad \text{and} \quad K \geq  \dfrac{M_TN}{TM_R}
\end{equation}
leading to the following bound on the required number of blocks
\begin{equation}
    K \geq \max\Big( \dfrac{M_TN}{I} ,  \dfrac{M_TN}{TM_R} \Big).
    \label{EQ:Necessray condition}
\end{equation}

It is important to note that the inequalities (\ref{EQ: condition to uniqueness for TALS-TUCKER}) and (\ref{EQ:Necessray condition}) are easy-to-check necessary conditions to ensure the uniqueness of the estimation of the joint channel and the symbol matrix. Additionally, they allow us to assess the degree of restriction imposed by the proposed solutions, thereby guiding the selection of appropriate system parameters for each semi-blind receiver and shedding light on the trade-offs between the two solutions. 
\begin{figure}[!t]
    \centering
    \includegraphics[scale = 0.57]{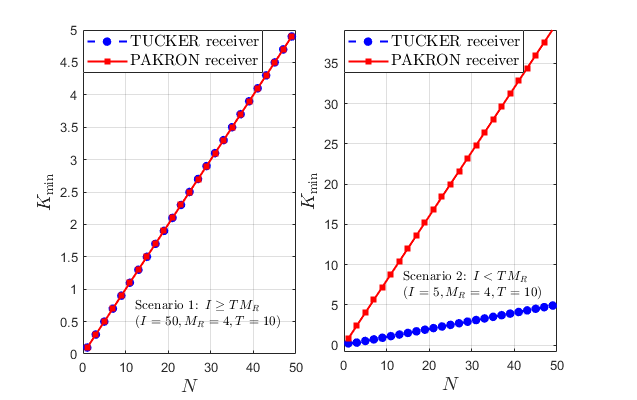}
    \caption{Minimum number $K$ of data blocks required for joint channel and symbol estimation for TUCKER and PAKRON semi-blind receivers, considering different \ac{BD-RIS} sizes.}
    \label{fig:K_restriction}
\end{figure}
To verify the design constraints of the proposed solutions, we keep $M_T$, $M_R$, and $T$ constant and determine $K_{\min}$ for two scenarios as a function of $N$. Figure \ref{fig:K_restriction} illustrates that, in the first scenario, where $I\geq TM_R$, both receivers have the same $K_{\min}$. In contrast, in the second scenario, when $I < TM_R$, the TUCKER receiver is considerably less restrictive compared to the PAKRON method. Additionally, stage I of the PAKRON receiver yields a sufficient condition for uniqueness. Since $\ten{Z}$ follows a third-order PARAFAC model, uniqueness is ensured if the so-called Kruskal condition is satisfied \cite{stegeman2007}
\begin{equation}
    Kr(\ma{\Omega}) + Kr(\ma{\Psi}) + Kr(\overline{\ma{G}}) \geq 2M_TN + 2.
    \label{EQ: Kruskal_rank}
\end{equation}
Therefore, it is important to analyze how these structures influence the uniqueness condition (\ref{EQ: Kruskal_rank}). To this end, note that $\ra(\ma{\Omega}) = \ra(\ma{X} \otimes \ma{HS}) = \ra(\ma{X})\ra(\ma{HS})$. Considering $T \geq M_T$, and using the fact the $\ma{S} \in \bb{C}^{N \times N}$ is full rank by construction, we have 
\begin{equation}
   \ra(\ma{\Omega}) = T\ra(\ma{H}) = T\min(M_R,N).
   \label{EQ: rank of Omega}
\end{equation}
 Recalling the lemma on the rank of the Khatri-Rao product (see \cite{SidNway2000,stegeman2007,LDL2008}), and using the fact that $\ma{W}$ and $\ma{P}$ are full-rank matrices by construction, $\ma{\Psi}$ has full rank equal to $K$. Additionally, assuming scattering-rich propagation, the Kruskal-rank condition (\ref{EQ: Kruskal_rank}) can be rewritten as a function of the system parameters as follows
\begin{equation}
\begin{split}
 K +  T\min(M_R,N) +  \min(I,M_TN) \geq 2M_TN + 2.
  \end{split}
  \label{EQ: Kruskal_Func_Pa}
\end{equation}
This condition sheds light on the balance of system parameters and establishes a sufficient condition for the PAKRON receiver to uniquely estimate the channel and symbols.

\textit{Remark 3:} It should be noted from (\ref{EQ: Kruskal_Func_Pa}) that various scenarios can be explored by varying the configurations of the system parameters. In this context, it is also worth noting that the PAKRON receiver still operates in a more challenging scenario under rank-deficient channel models, as it may not satisfy the condition (\ref{EQ: Kruskal_Func_Pa}). For example, in the extreme case where the rank of the RIS-BS channel $\ma{H}$ is one (which arises under line-of-sight (LoS) propagation between the BS and the RIS), we have from (\ref{EQ: rank of Omega}) that  $\ra(\ma{\Omega}) = T$. In this case, when we collect a sufficient number of frames so that $I \geq M_TN$, condition (\ref{EQ: Kruskal_Func_Pa}) is automatically fulfilled. Otherwise, if $I < M_TN$, it becomes necessary to have $T + K + I \geq 2M_TN + 2$ to compensate for the rank deficiency of the channels.

\textit{Complexity:} In algorithms involving pseudo-inverse calculations or rank-$1$ approximations, the computational complexity is generally dominated by the Singular Value Decomposition (SVD) \cite{Favier2019}. Note that for an arbitrary matrix $\ma{A} \in \bb{C}^{M \times N}$, the complexity of its SVD computation is $\ten{O}(MN\min(M,N))$. It can be seen that the complexity of solving an \ac{LS} problem is dominated by the associated SVD computations. Considering that the PAKRON algorithm comprises two stages, we have: In Stage I, the computational complexity is primarily determined by steps $1$ and $2$, which have complexities of $\ten{O}((M_TN)^2IK)$ and $\ten{O}((M_TN)^2KTM_R)$, respectively. In Stage II, the dominant computation corresponds to a single rank-$1$ approximation, which has complexity $\ten{O}(M_TNTM_R)$. Then the asymptotic per-iteration complexity of the PAKRON receiver is $\max\big((M_TN)^2IK, (M_TN)^2KTM_R, M_TNTM_R\big)$. On the other hand, the computational complexity of the TUCKER receiver is dictated by steps $1$, $2$, and $3$, which have complexities $\ten{O}((N)^2IKT)$, $\ten{O}(IKM_R(M_T)^2)$, and $\ten{O}(KTM_R(M_TN)^2)$, respectively. Therefore, the asymptotical complexity of the TUCKER receiver is $\max\big((N)^2IKT, IKM_R(M_T)^2, KTM_R(M_TN)^2\big)$. Table~\ref{tab:complexity table} summarizes the asymptotic complexities of the identifiability conditions of our proposed receivers compared to competing methods. It is important to emphasize that BALS (\cite{Sokal_BD_RIS_2024}) requires the simplest receiver and consequently generally exhibits the lowest computational complexity in terms of flops compared to the TUCKER Kronecker factorization BTKF,\footnote{Block Tucker Kronecker factorization (BTKF).} PAKRON and TUCKER receivers. Moreover, as the number of \ac{BD-RIS} scattering elements increases, the TUCKER receiver becomes less complex than the BTKF receiver, which in turn is less complex than PAKRON (although this gap noticeably narrows as $N$ grows). Among all the schemes considered, PAKRON usually incurs the highest flop count. Note, however, that the higher computational complexity of our proposed methods is offset by their ability to operate semi-blindly, without requiring pilot sequences. 

\begin{table*}[t!]  
\centering
\begingroup
\small
\setlength{\tabcolsep}{12pt}
\renewcommand{\arraystretch}{1.5}

\begin{tabular}{l c c r}
\toprule
\thead{Receiver} & \thead{Complexity} & \thead{Identifiability condition} &  \thead{Approach} \\
\midrule

PAKRON
& $\max\!\Bigl(
   (M_TN)^2IK\allowbreak,\;
   (M_TN)^2KTM_R\allowbreak,\;
   M_TNTM_R\Bigr)$
& $ K \ge \max\!\Bigl(
   \frac{M_TN}{I}\allowbreak,\,
   \frac{M_TN}{TM_R}\Bigr)$
& Semi-Blind \\

TUCKER
& $ \max\!\Bigl(
   N^2IKT\allowbreak,\;
   IKM_R(M_T)^2\allowbreak,\;
   KTM_R(M_TN)^2\Bigr)$
& $ K \ge \max\!\Bigl(
   \frac{N}{IT}\allowbreak,\,
   \frac{M_T}{IM_R}\allowbreak,\,
   \frac{M_TN}{TM_R}\Bigr)$
& Semi-Blind \\

LS
& $ \ten{O}\bigl(K(\frac{N^2}{Q})^2\bigr)$ \cite{B_Clerckc_CE_TSP_24}
& $ K \ge \frac{N^2}{Q}$
& Pilot-Assisted \\

LS-KRF
& $ \ten{O}\bigl(K(\frac{N^2}{Q})^2 + M_RM_T\frac{N^2}{Q}\bigr)$ \cite{Sokal_Asilomar_2024}
& $ K \ge \frac{N^2}{Q}$
& Pilot-Assisted \\

BTKF
& $ \max\!\Bigl(
   M_RM_T\dfrac{N^2}{Q}\allowbreak,\;
   K(\dfrac{N^2}{Q})^2\Bigr)$ \cite{Sokal_BD_RIS_2024}
& $ K \ge \dfrac{N^2}{Q}$
& Pilot-Assisted \\

BTALS
& $\max\!\Bigl(
   KM_T(\dfrac{N^2}{Q})^2\allowbreak,\;
   KM_R(\dfrac{N^2}{Q})^2\Bigr)$ \cite{Sokal_BD_RIS_2024}
& $ K \ge \max\!\Bigl(
   \frac{N}{M_T}\allowbreak,\,
   \frac{N}{M_R}\Bigr)$
& Pilot-Assisted \\
\bottomrule
\end{tabular}
\endgroup
\caption{Summary of identifiability conditions and computational complexity of main solutions in the literature.}
\label{tab:complexity table}
\end{table*}

\section{SIMULATION RESULTS}
In this section, we present several numerical experiments to evaluate the performance of our proposed semi-blind receivers under various setups. The channel estimation accuracy is evaluated in terms of \ac{NMSE}, which is given by 
\begin{equation}
    \textrm{NMSE}(\hat{\ma{\Xi}^{(r)}}) = \frac{1}{R}\sum_{r=1}^{R} \dfrac{\|\ma{\Xi}^{(r)} - \hat{\ma{\Xi}}^{(r)}\|_F^2}{ \|\ma{\Xi}^{(r)}\|_F^2},
    \label{EQ:NMSE}
\end{equation}
 where $\ma{\Xi} \in \{\ma{H},\overline{\ma{G}}\}$ and $\hat{\ma{\Xi}}^{(r)}$ is their estimate obtained in the $r$ -th Monte Carlo run. We also show numerical results in terms of the \ac{SER} performance as a function of the \ac{SNR} in dB. For comparison, we also plot the performance of competing (pilot-assisted) baseline methods, such as the LS-based channel estimation method proposed in \cite{Clerckx_CAMSAP_2023}, and the BTKF and BTALS methods proposed recently in \cite{Sokal_BD_RIS_2024}. The performance of the traditional zero-forcing receiver, which operates with perfect channel knowledge, is also plotted. The results are averaged over $R = 1000$ Monte Carlo runs. Regarding the channel model, we consider the Rayleigh fading case (i.e., the entries of the channel matrices are independent and identically distributed zero-mean circularly symmetric complex Gaussian random variables), as well as a multipath (low-rank) geometric channel model, in which the path directions are randomly generated according to a uniform distribution. In this case, at each Monte Carlo run, the azimuth and elevation angles are drawn within the intervals $[-\pi/2, \pi/2]$ and $[0, \pi/2]$, respectively.  The system parameters used in this analysis were \(N = 16\), \(M_R = 4\), \(M_T = 2\), \(I = 2\), \(K = 32\), \(T = 4\), and \(Q \in \{2, 4, 8\}\). The system parameters were selected to ensure that the identifiability conditions for all compared methods were satisfied simultaneously, ensuring a fair comparison under feasible operating conditions.

\begin{figure}[!t]
    \centering
    \includegraphics[scale = 0.59]{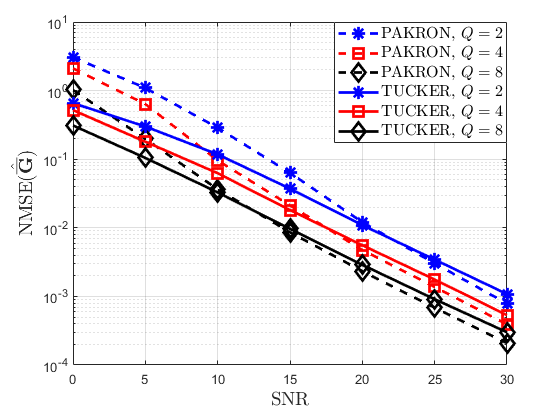}
    \caption{NMSE performance of the user to the \ac{BD-RIS} channels.}
    \label{fig:G_Edited}
\end{figure}
\begin{figure}[!t]
    \centering
    \includegraphics[scale = 0.59]{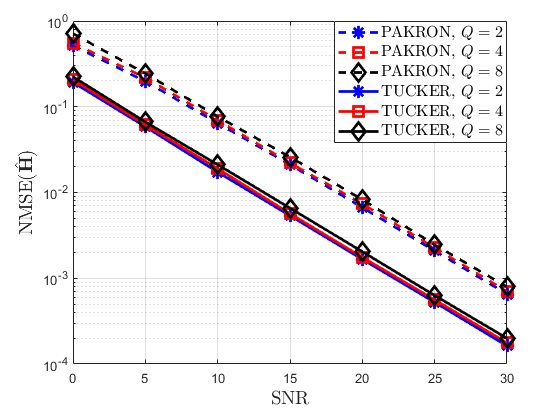}
    \caption{NMSE performance of the  the \ac{BD-RIS} to BS channels.}
    \label{fig:H_Edited}
\end{figure}
\begin{figure}
    \centering
    \includegraphics[scale = 0.59]{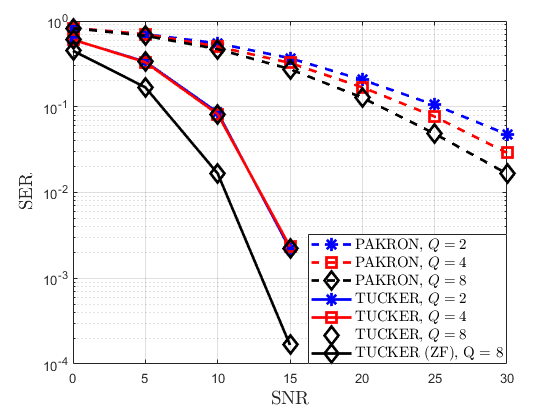}
    \caption{SER performance of the proposed algorithm.}
    \label{fig:SER}
\end{figure}
Figures \ref{fig:G_Edited} and \ref{fig:H_Edited} illustrate the channel estimation accuracy of the PAKRON and TUCKER receivers. Let us first consider Figure \ref{fig:G_Edited}, which depicts the performance of the estimated user on the BD-RIS channel (\(\overline{\ma{G}}\)). It should be noted that the PAKRON receiver exhibits greater degradation than the TUCKER receiver at lower SNR levels (in particular, up to $ 10$ dB). However, in the higher-SNR regime, this performance gap tends to diminish, yielding nearly identical performance, with the PAKRON receiver showing a slight tendency to outperform the TUCKER receiver. Moreover, we observe that both algorithms are sensitive to variations in the number of groups \(Q\). This sensitivity is anticipated, as the number of blocks can influence the iterative part of the PAKRON receiver, specifically the estimation of the channel \(\overline{\ma{G}}\) and the structured matrix \(\ma{\Omega}\). On the other hand, Figure \ref{fig:H_Edited} shows that the TUCKER receiver consistently outperforms the PAKRON receiver throughout the entire SNR range. Nevertheless, both solutions show insensitivity to the number of groups. For the PAKRON receiver, the estimate of the channel $\ma{H}$ is extracted by applying a Kronecker factorization to the matrix $\ma{\Omega}$ (stage II) already estimated during the iterative part (stage I). This process reduces the sensitivity of the estimation of the \(\ma{H}\) channel with respect to variations in the number of groups. However, the propagation of errors from stage I to stage II explains the gap between the PAKRON and TUCKER receivers, favoring the latter. In summary, PAKRON operates on a third-order tensor model whose factor matrices are constrained. This induces a coupling between the dimensions of the original tensor. As a consequence, it becomes more sensitive to changes in the number of groups and suffers performance degradation in channel estimation $\ma{G}$ at low SNR compared to the TUCKER receiver. The TUCKER receiver, in contrast, preserves the fourth-order tensor structure of the received signal and thus does not incur diversity loss, since no dimension coupling occurs. As a result, it achieves more accurate estimates of $\ma{G}$ in the low SNR regime. Nevertheless, this advantage tends to diminish as $Q$ grows. For channel $\ma{H}$, the TUCKER receiver also exceeds the PAKRON receiver in all simulated SNR ranges. This is mainly because it retains the diversity offered by the fourth-order tensor structure while enabling single-stage processing, thereby avoiding the extra stage required by PAKRON. This additional stage leads to error propagation, which appears to be the primary reason for the performance disparity between the two receivers in estimating the channel matrix $\ma{H}$.

The impact of this error propagation is also evident in Figure \ref{fig:SER}, where the TUCKER receiver significantly outperforms the PAKRON receiver, particularly with a higher-order $64$-PSK constellation scheme. We also compare the TUCKER receiver with the clairvoyant \ac{ZF} receiver derived directly from (\ref{EQ:2-mode tucker}), which assumes perfect channel knowledge. \footnote{It is worth emphasizing that the proposed semi-blind receivers are not intended to replace pilot-assisted channel estimation schemes in all operating conditions. Instead, they provide an alternative trade-off between channel estimation accuracy, training overhead, latency, and spectral efficiency. By exploiting useful data symbols for channel acquisition, the proposed framework eliminates the need for a dedicated pilot transmission phase, thereby improving spectral efficiency and reducing decoding latency.} The results indicate strong performance compared with the ideal-channel scenario. Additionally, the performance difference stems from the fact that, for the PAKRON receiver, the received signal tensor is reduced from fourth order to third order. Due to this order reduction, the original tensor structure is lost, leading to mode mixing. This loss prevents the tensor structure from being fully exploited, a capability that the TUCKER receiver retains by maintaining the original tensor order.
 
\begin{figure}[!t]
    \centering
    \includegraphics[scale = 0.59]{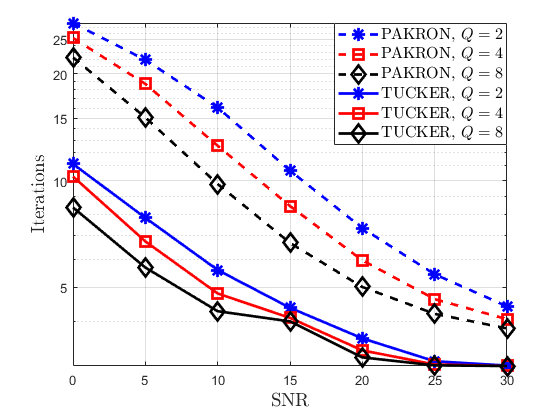}
    \caption{Average number of iterations.}
    \label{fig:IT}
\end{figure}

\begin{figure}[!t]
    \centering
    \includegraphics[scale = 0.59]{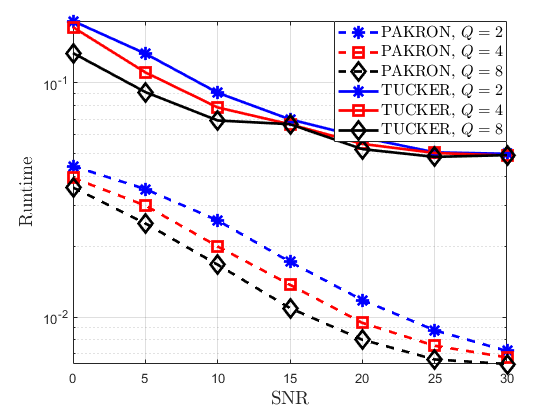}
    \caption{Average run time}
    \label{fig:RT}
\end{figure}

The complexity of the proposed solution is evaluated using two key metrics: the average number of iterations required for convergence and the corresponding average run time. These metrics are shown in Figs.~\ref{fig:IT} and \ref{fig:RT}. Notably, while the PAKRON algorithm requires more iterations to converge than the TUCKER algorithm, it achieves a shorter execution time. This implies that PAKRON can complete more iterations within the same time interval, indicating a more efficient use of computational resources. Consequently, the TUCKER algorithm, despite having fewer iterations, is considered computationally more complex.  This difference arises from the distinct computational demands of the two algorithms. Specifically, PAKRON computes two pseudo-inverses, whereas TUCKER requires three. Given the high computational cost of computing pseudo-inverses, especially as the dimensions of the problem matrices increase, the TUCKER approach incurs a higher computational cost. 

\begin{figure}
    \centering
    \includegraphics[scale = 0.47]{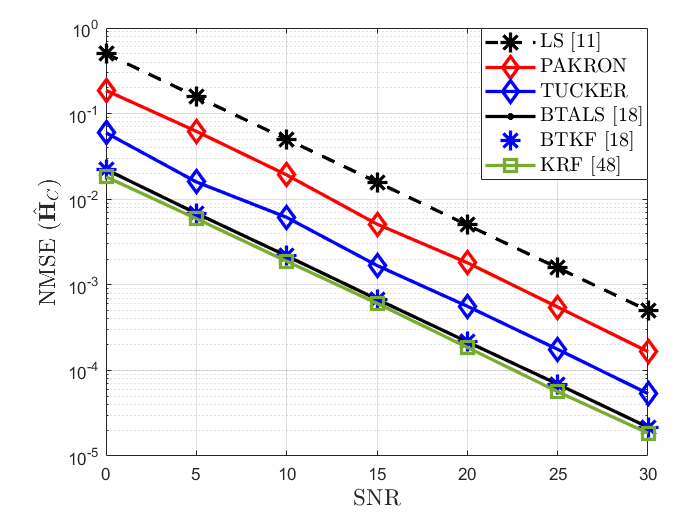}
    \caption{NMSE performance of the composite channel compared with existing methods.}
    \label{fig:comp_Bruno}
\end{figure}

Figure \ref{fig:comp_Bruno} compares the proposed algorithms with established methods from the literature, namely the method in \cite{B_Clerckc_CE_TSP_24} and the tensor-based approach described in \cite{Sokal_Asilomar_2024} for $Q = 2$. The work in \cite{B_Clerckc_CE_TSP_24} provides a conventional \ac{LS} solution, while \cite{Sokal_Asilomar_2024} and \cite{Sokal_BD_RIS_2024} introduce tensor-based approaches, namely LS-KRF \cite{Sokal_Asilomar_2024}, BTKF \cite{Sokal_BD_RIS_2024}, and BTALS \cite{Sokal_BD_RIS_2024}, for composite channel estimation. The composite channel is characterized as a block-Kronecker product of the involved channels, expressed as $\ma{H}_C = \ma{H} |\otimes| \ma{G} = [\ma{H}^{(1)} |\otimes| \ma{G}^{(1)} \dots \ma{H}^{(Q)} |\otimes| \ma{G}^{(Q)}]$. Unlike our approach, which assumes that the channel $\ma{G}$ varies dynamically over time, the competing methods assume a quasi-static channel. To ensure a fair and rigorous comparison, the \ac{NMSE} of the proposed composite channel estimate is calculated by averaging across $I$ composite channels, where $\ma{H}_C^i = \ma{H} |\otimes| \ma{G}_i$ for \(i = 1, \dots, I\). Considering the setup $M_R = 10$, $M_T = 2$, $K = 32$, $I = 4$, $T = 8$, $N = 16$, $Q=4$, and an i.i.d. channel model, the proposed TUCKER and PAKRON receivers achieve good performance and significantly outperform the \ac{LS} method described in \cite{B_Clerckc_CE_TSP_24}. This outcome is expected, given that the simplicity of the LS solution does not exploit the intrinsic tensor structure of the signal model. In contrast, our approach leverages this structure to achieve noise-rejection gain. Moreover, although the competing receivers analyzed in \cite{Sokal_Asilomar_2024} and \cite{Sokal_BD_RIS_2024} achieve superior results, this behavior is expected because both the proposed and competing algorithms benefit from effective noise rejection. However, our semi-blind receivers require the estimation of three matrices ($\ma{H}$, $\overline{\ma{G}}$, and $\ma{X}$), compared to the LS-KRF, BTALS, and BTKF receivers, which are limited to estimating the first two. Nevertheless, the proposed semi-blind methods also provide early estimates of the data symbols, unlike competing methods that lack this feature and require a dedicated pilot transmission period prior to symbol decoding. 

\begin{figure}[!ht]
    \centering
    \includegraphics[scale = 0.59]{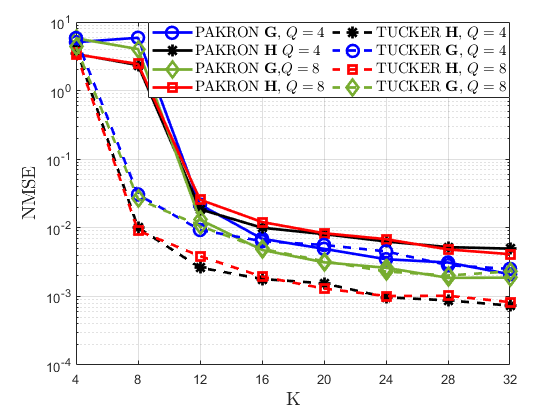}
    \caption{NMSE performance for different values of blocks $K$}
    \label{fig:NMSE_K}
\end{figure}

From a fundamental standpoint, the parameter $K$ must at least satisfy the lower bounds imposed by the identifiability conditions. In practical implementations, however, this minimal admissible value of $K$ may be insufficient to guarantee satisfactory receiver performance. Increasing $K$ generally improves the NMSE. Nonetheless, as illustrated in Figure \ref{fig:NMSE_K}, choosing an excessively large $K$ does not result in unbounded gains in NMSE reduction. As $K$ increases, the system performance tends to converge to a steady state, beyond which further increments in $K$ become suboptimal. This behavior arises from the inherent trade-off between spectral efficiency and estimation accuracy.

\begin{figure}[!ht]
    \centering
    \includegraphics[scale = 0.59]{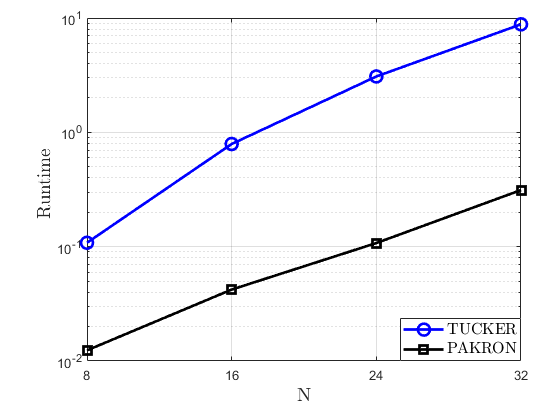}
    \caption{Runtime versus $N$ to illustrate the scalability of the proposed algorithm}
    \label{fig:RT_N}
\end{figure}

 Figure \ref{fig:RT_N} illustrates the computational scalability of the proposed receivers by plotting the average run time as a function of the number of BD-RIS elements $N$. As expected, the run time for both receivers grows with increasing $N$, which aligns with the complexity assessment in Table I. There, it is shown that the dominant cost in each ALS iteration stems from pseudo-inverse computations on matrices whose dimensions scale with $M_T N$. This occurs because the channel matrices become larger, increasing the number of coefficients to be estimated and, consequently, slowing the convergence of the algorithms.

\textit{Remark 4:} The results are influenced by the designs of the matrices $\ma{P}$ and $\ma{W}$, and consequently by $\ma{\Psi}$. We focus on the design described in Section \ref{Sec: BD-RIS Design}. It is essential to note that the optimal design of the rotation matrix $\mathbf{P}$ and the coding matrix $\mathbf{W}$ can significantly affect the results. Exploring this aspect is beyond the scope of this paper.

\section{CONCLUSION}
In this paper, we proposed a novel tensor-based semi-blind receiver design for \ac{BD-RIS}-assisted \ac{MIMO} communication systems with a time-varying \ac{UT}-\ac{BD-RIS} channel, which yields a fourth-order PARATUCK tensor model of the received signals. Based on this model, we formulated data-aided semi-blind receivers that avoid dedicated pilot sequences while jointly estimating the \ac{BD-RIS}-\ac{BS} channel, the \ac{UT}-\ac{BD-RIS} channel, and the transmitted symbols. Specifically, two receivers were proposed. The PAKRON receiver (Algorithm~\ref{Algo:PAKRON}) is a two-stage approach in which the data symbols are first estimated jointly with a structured matrix $\ma{\Omega}$, which is composed of the Kronecker product between the \ac{BD-RIS}-\ac{BS} channel and the data-symbol matrix through an iterative \ac{BALS} procedure. In the second stage, Kronecker factorization is applied to the estimated version of $\ma{\Omega}$ to estimate the \ac{BD-RIS}-\ac{BS} channel component and the symbols. The TUCKER receiver is a fourth-order tensor-decomposition method in which the channels and data are jointly and iteratively estimated in a single stage using the \ac{TALS} algorithm (Algorithm~\ref{Algo: Pseudocode TALS-TUCKER}). 

Our results indicate that the TUCKER receiver is more efficient than the PAKRON receiver for channel estimation and data symbol accuracy. However, it is more computationally complex than PAKRON. Additionally, the two proposed methods outperform the competing \ac{LS} method, which does not exploit the inherent multidimensional structure of the received signal. Recall that the competing LS-KRF method, used as a reference, is a pilot-assisted scheme, whereas our approach is a semi-blind scheme suitable for scenarios with limited or unavailable pilot resources. A possible hybrid implementation of the proposed semi-blind receiver uses the soft estimates of the data symbol matrix from PAKRON as initialization for the TUCKER receiver. Another perspective of this work is to investigate the optimal design of the matrix $\ma{\Psi}$. Additionally, new tensor structures and the Cram\'er-Rao bound can be derived for the different receivers. An immediate avenue for future research arising from this work is to investigate the proposed semi-blind approach under practical hardware-impairment conditions, explicitly accounting for inter-group leakage and coupling effects in the BD-RIS architecture.

\bibliographystyle{IEEEtran}
\bibliography{ref.bib}

\begin{thebibliography}{10}
\providecommand{\url}[1]{#1}
\csname url@samestyle\endcsname
\providecommand{\newblock}{\relax}
\providecommand{\bibinfo}[2]{#2}
\providecommand{\BIBentrySTDinterwordspacing}{\spaceskip=0pt\relax}
\providecommand{\BIBentryALTinterwordstretchfactor}{4}
\providecommand{\BIBentryALTinterwordspacing}{\spaceskip=\fontdimen2\font plus
\BIBentryALTinterwordstretchfactor\fontdimen3\font minus
  \fontdimen4\font\relax}
\providecommand{\BIBforeignlanguage}[2]{{%
\expandafter\ifx\csname l@#1\endcsname\relax
\typeout{** WARNING: IEEEtran.bst: No hyphenation pattern has been}%
\typeout{** loaded for the language `#1'. Using the pattern for}%
\typeout{** the default language instead.}%
\else
\language=\csname l@#1\endcsname
\fi
#2}}
\providecommand{\BIBdecl}{\relax}
\BIBdecl

\bibitem{RIS_6G_2021}
C.~Pan, H.~Ren, K.~Wang, J.~F. Kolb, M.~Elkashlan, M.~Chen, M.~Di~Renzo,
  Y.~Hao, J.~Wang, A.~L. Swindlehurst, X.~You, and L.~Hanzo, ``Reconfigurable
  intelligent surfaces for {6G} systems: Principles, applications, and research
  directions,'' \emph{IEEE Commun. Mag.}, vol.~59, no.~6, pp. 14--20, 2021.

\bibitem{RIS_Application}
S.~Basharat, S.~A. Hassan, H.~Pervaiz, A.~Mahmood, Z.~Ding, and M.~Gidlund,
  ``Reconfigurable intelligent surfaces: Potentials, applications, and
  challenges for {6G} wireless networks,'' \emph{IEEE Wireless Commun.},
  vol.~28, no.~6, pp. 184--191, 2021.

\bibitem{RIS_Opotunit}
X.~Yuan, Y.-J.~A. Zhang, Y.~Shi, W.~Yan, and H.~Liu,
  ``Reconfigurable-intelligent-surface empowered wireless communications:
  Challenges and opportunities,'' \emph{IEEE Wireless Commun.}, vol.~28, no.~2,
  pp. 136--143, 2021.

\bibitem{Hibrid_Alexandropoulos_2023_MAG}
G.~C. Alexandropoulos, N.~Shlezinger, I.~Alamzadeh, M.~F. Imani, H.~Zhang, and
  Y.~C. Eldar, ``Hybrid reconfigurable intelligent metasurfaces: Enabling
  simultaneous tunable reflections and sensing for {6G} wireless
  communications,'' \emph{IEEE Veh. Technol. Mag.}, vol.~19, no.~1, pp. 75--84,
  2024.

\bibitem{Hibrid_Alexandropoulos_2023_TCOM}
H.~Zhang, N.~Shlezinger, G.~C. Alexandropoulos, A.~Shultzman, I.~Alamzadeh,
  M.~F. Imani, and Y.~C. Eldar, ``Channel estimation with hybrid reconfigurable
  intelligent metasurfaces,'' \emph{IEEE Trans. Commun.}, vol.~71, no.~4, pp.
  2441--2456, 2023.

\bibitem{Amarilton_2024_ISWCS}
A.~L. Magalhães and A.~L.~F. de~Almeida, ``Joint channel and symbol estimation
  for hybrid {RIS} wireless communications,'' in \emph{(ISWCS)}, 2024, pp.
  1--6.

\bibitem{STAR_2023_JSTSP}
Z.~Wang, X.~Mu, J.~Xu, and Y.~Liu, ``Simultaneously transmitting and reflecting
  surface ({STARS}) for terahertz communications,'' \emph{IEEE J. Sel. Topics
  Signal Process.}, vol.~17, no.~4, pp. 861--877, 2023.

\bibitem{STAR_2023_TCOM}
X.~Zhai, G.~Han, Y.~Cai, Y.~Liu, and L.~Hanzo, ``Simultaneously transmitting
  and reflecting ({STAR}) {RIS} assisted over-the-air computation systems,''
  \emph{IEEE Trans. Commun.}, vol.~71, no.~3, pp. 1309--1322, 2023.

\bibitem{STAR_RIS_360_mag}
Y.~Liu, X.~Mu, J.~Xu, R.~Schober, Y.~Hao, H.~V. Poor, and L.~Hanzo, ``{STAR:
  Simultaneous Transmission and Reflection for 360º; Coverage by Intelligent
  Surfaces},'' \emph{IEEE Wireless Commun.}, vol.~28, no.~6, pp. 102--109,
  2021.

\bibitem{Clerckx_TWC_APR_2023}
H.~Li, S.~Shen, and B.~Clerckx, ``Beyond diagonal reconfigurable intelligent
  surfaces: From transmitting and reflecting modes to single-, group-, and
  fully-connected architectures,'' \emph{IEEE Trans. Wireless Commun.},
  vol.~22, no.~4, pp. 2311--2324, 2023.

\bibitem{B_Clerckc_CE_TSP_24}
H.~Li, S.~Shen, Y.~Zhang, and B.~Clerckx, ``Channel estimation and beamforming
  for beyond diagonal reconfigurable intelligent surfaces,'' \emph{IEEE Trans.
  Signal Process.}, vol.~72, pp. 3318--3332, 2024.

\bibitem{Clerckx_TVT_JUN_2022}
Q.~Li, M.~El-Hajjar, I.~Hemadeh, A.~Shojaeifard, A.~A.~M. Mourad, B.~Clerckx,
  and L.~Hanzo, ``Reconfigurable intelligent surfaces relying on non-diagonal
  phase shift matrices,'' \emph{IEEE Trans. Vehicular Tech.}, vol.~71, no.~6,
  pp. 6367--6383, 2022.

\bibitem{Shen_2022}
S.~Shen, B.~Clerckx, and R.~Murch, ``Modeling and architecture design of
  reconfigurable intelligent surfaces using scattering parameter network
  analysis,'' \emph{IEEE Trans. Wireless Commun.}, vol.~21, no.~2, pp.
  1229--1243, 2022.

\bibitem{matteo_2024}
M.~Nerini, S.~Shen, H.~Li, M.~Di~Renzo, and B.~Clerckx, ``A universal framework
  for multiport network analysis of reconfigurable intelligent surfaces,''
  \emph{IEEE Trans. Wireless Commun.}, vol.~23, no.~10, pp. 14\,575--14\,590,
  2024.

\bibitem{Clerckx_TWC_FEB_2024}
M.~Nerini, S.~Shen, and B.~Clerckx, ``Closed-form global optimization of beyond
  diagonal reconfigurable intelligent surfaces,'' \emph{IEEE Trans. Wireless
  Commun.}, vol.~23, no.~2, pp. 1037--1051, 2024.

\bibitem{B.Clerckx_TWC_EA_2024}
M.~Soleymani, I.~Santamaria, E.~Jorswieck, and B.~Clerckx, ``Optimization of
  rate-splitting multiple access in beyond diagonal {RIS}-assisted urllc
  systems,'' \emph{IEEE Trans. Wireless Commun.}, pp. 1--1, 2023.

\bibitem{zheng_survey}
B.~Zheng, C.~You, W.~Mei, and R.~Zhang, ``A survey on channel estimation and
  practical passive beamforming design for intelligent reflecting surface aided
  wireless communications,'' \emph{IEEE Commun. Surveys Tuts.}, vol.~24, no.~2,
  pp. 1035--1071, Feb. 2022.

\bibitem{Sokal_BD_RIS_2024}
A.~L.~F. de~Almeida, B.~Sokal, H.~Li, and B.~Clerckx, ``Channel estimation for
  beyond diagonal {RIS} via tensor decomposition,'' \emph{IEEE Trans. Signal
  Process.}, 2025.

\bibitem{Gil_Asilomar}
G.~T. de~Araújo and A.~L.~F. de~Almeida, ``Semi-blind channel estimation for
  beyond diagonal {RIS},'' in \emph{(ASILOMAR)}, 2024, pp. 1586--1590.

\bibitem{Gil_TSP}
G.~T. de~Araújo, A.~L.~F. de~Almeida, R.~Boyer, and G.~Fodor, ``Semi-blind
  joint channel and symbol estimation for {IRS}-assisted {MIMO} systems,''
  \emph{IEEE Trans. Sig. Proces.}, vol.~71, pp. 1184--1199, 2023.

\bibitem{Suggestion_RI}
J.~V. Alegría, J.~Thunberg, and O.~Edfors, ``Channel orthogonalization with
  reconfigurable surfaces: General models, theoretical limits, and effective
  configuration,'' \emph{IEEE Trans. Wireless Commun.}, vol.~24, no.~6, pp.
  5179--5195, 2025.

\bibitem{Sun2024PowerMeasurementIRS}
H.~Sun, L.~Zhu, W.~Mei, and R.~Zhang, ``Power measurement-based channel
  estimation for {IRS}-enhanced wireless coverage,'' \emph{IEEE Trans. Wireless
  Commun.}, vol.~23, no.~12, pp. 19\,183--19\,198, 2024.

\bibitem{Sun2025PowerAutocorrelationIRS}
------, ``Power-measurement-based channel autocorrelation estimation for
  {IRS}-assisted wideband communications,'' \emph{IEEE Trans. Wireless
  Commun.}, vol.~24, no.~6, pp. 4647--4662, 2025.

\bibitem{Liu2025PowerMeasurementBDRIS}
Y.~Liu, W.~Mei, H.~Sun, D.~Wang, and Z.~Chen, ``Power-measurement-based channel
  estimation for beyond diagonal {RIS},'' \emph{IEEE Communications Letters},
  vol.~29, no.~11, pp. 2666--2670, 2025.

\bibitem{Almeida_Elsevier_2007}
A.~L.~F. {de Almeida}, G.~Favier, and J.~C.~M. Mota, ``{PARAFAC}-based unified
  tensor modeling for wireless communication systems with application to blind
  multiuser equalization,'' \emph{Signal Processing}, vol.~87, no.~2, pp.
  337--351, 2007, tensor Signal Processing.

\bibitem{Almeida2008}
A.~L.~F. de~Almeida, G.~Favier, and J.~C.~M. Mota, ``A constrained factor
  decomposition with application to mimo antenna systems,'' \emph{IEEE Trans.
  Signal Process.}, vol.~56, no.~6, pp. 2429--2442, 2008.

\bibitem{Favier2014}
G.~Favier and A.~L.~F. de~Almeida, ``Tensor space-time-frequency coding with
  semi-blind receivers for {MIMO} wireless communication systems,'' \emph{IEEE
  Trans. Signal Process.}, vol.~62, no.~22, pp. 5987--6002, 11 2014.

\bibitem{Ximenes_2015}
L.~R. Ximenes, G.~Favier, and A.~L.~F. de~Almeida, ``Semi-blind receivers for
  non-regenerative cooperative {MIMO} communications based on nested {PARAFAC}
  modeling,'' \emph{IEEE Trans. Signal Process.}, vol.~63, no.~18, pp.
  4985--4998, 2015.

\bibitem{Ximenes_2014}
L.~R. Ximenes, G.~Favier, A.~L.~F. de~Almeida, and Y.~C.~B. Silva,
  ``{PARAFAC-PARATUCK} semi-blind receivers for two-hop cooperative {MIMO}
  relay systems,'' \emph{IEEE Trans. Sig. Proces.}, vol.~62, no.~14, pp.
  3604--3615, 2014.

\bibitem{Chen2021}
H.~Chen, F.~Ahmad, S.~Vorobyov, and F.~Porikli, ``Tensor decompositions in
  wireless communications and {MIMO} radar,'' \emph{IEEE J. Sel. Topics Signal
  Process.}, vol.~15, no.~3, pp. 438--453, 2021.

\bibitem{Fazal_2024_TVT}
Fazal-E-Asim, B.~Sokal, A.~L.~F. de~Almeida, B.~Makki, and G.~Fodor,
  ``Structured channel estimation for {RIS}-assisted {THz} communications,''
  \emph{IEEE Trans. Veh. Technol.}, pp. 1--6, 2024.

\bibitem{Fazal_2024_TCOM}
Fazal-E-Asim, A.~L.~F. De~Almeida, B.~Sokal, B.~Makki, and G.~Fodor,
  ``Two-dimensional channel parameter estimation for {IRS}-assisted networks,''
  \emph{IEEE Trans. Commun.}, pp. 1--1, 2024.

\bibitem{Yuri_Sales_2024_WCL}
Y.~S. Ribeiro, A.~L.~F. de~Almeida, Fazal-E-Asim, B.~Makki, and G.~Fodor,
  ``Low-complexity joint active and passive beamforming design for
  {IRS}-assisted {MIMO},'' \emph{IEEE Wireless Commun. Lett.}, vol.~13, no.~3,
  pp. 607--611, 2024.

\bibitem{Sokal_2023_TWC}
B.~Sokal, P.~R.~B. Gomes, A.~L.~F. de~Almeida, B.~Makki, and G.~Fodor,
  ``Reducing the control overhead of intelligent reconfigurable surfaces via a
  tensor-based low-rank factorization approach,'' \emph{IEEE Trans. Wireless
  Commun.}, vol.~22, no.~10, pp. 6578--6593, 2023.

\bibitem{Gil_JTSP}
G.~T. {de Araújo}, A.~L.~F. {de Almeida}, and R.~{Boyer}, ``Channel estimation
  for intelligent reflecting surface assisted {MIMO} systems: A tensor modeling
  approach,'' \emph{IEEE J. Sel. Topics Signal Process.}, vol.~15, no.~3, pp.
  789--802, Apr 2021.

\bibitem{paulo2022tensor}
P.~R.~B. Gomes, G.~T.~d. Araújo, B.~Sokal, A.~L. F.~d. Almeida, B.~Makki, and
  G.~Fodor, ``{Channel estimation in {RIS}-assisted MIMO systems operating
  under imperfections},'' \emph{IEEE Trans. Veh. Technol.}, pp. 1--14, 2023.

\bibitem{Gherekhloo23}
S.~Gherekhloo, K.~Ardah, A.~L.~F. De~Almeida, M.~Maleki, and M.~Haardt,
  ``Nested {PARAFAC} tensor-based channel estimation method for double
  {RIS}-aided {MIMO} communication systems,'' in \emph{(EUSIPCO)}, 2023, pp.
  1674--1678.

\bibitem{Nwalozie24}
G.~C. Nwalozie, A.~L. {de Almeida}, and M.~Haardt, ``Enhanced channel
  estimation for double {RIS}-aided {MIMO} systems using coupled tensor
  decompositions,'' \emph{Signal Processing}, vol. 234, p. 109979, 2025.

\bibitem{stacked}
Q.~Li, M.~El-Hajjar, C.~Xu, J.~An, C.~Yuen, and L.~Hanzo, ``Stacked intelligent
  metasurfaces for holographic {MIMO}-aided cell-free networks,'' \emph{IEEE
  Transactions on Communications}, vol.~72, no.~11, pp. 7139--7151, 2024.

\bibitem{holographic}
Q.~Li, M.~El-Hajjar, K.~Cao, C.~Xu, H.~Haas, and L.~Hanzo, ``Holographic
  metasurface-based beamforming for multi-altitude leo satellite networks,''
  \emph{IEEE Trans. Wireless Commun.}, vol.~24, no.~4, pp. 3103--3116, 2025.

\bibitem{Performance}
Q.~Li, M.~El-Hajjar, Y.~Sun, and L.~Hanzo, ``Performance analysis of
  reconfigurable holographic surfaces in the near-field scenario of cell-free
  networks under hardware impairments,'' \emph{IEEE Trans. Wireless Commun.},
  vol.~23, no.~9, pp. 11\,972--11\,984, 2024.

\bibitem{Sidiropoulos2017}
N.~D. Sidiropoulos, L.~De~Lathauwer, X.~Fu, K.~Huang, E.~E. Papalexakis, and
  C.~Faloutsos, ``Tensor decomposition for signal processing and machine
  learning,'' \emph{IEEE Transactions on Signal Processing}, vol.~65, no.~13,
  pp. 3551--3582, 2017.

\bibitem{Kolda2009}
T.~G. Kolda and B.~W. Bader, ``Tensor decompositions and applications,''
  \emph{SIAM Review}, vol.~51, no.~3, pp. 455--500, Aug. 2009.

\bibitem{favier2012tensor}
G.~Favier, M.~N. da~Costa, A.~L.~F. de~Almeida, and J.~M.~T. Romano, ``{Tensor
  space--time {(TST)} coding for {MIMO} wireless communication systems},''
  \emph{Signal Processing}, vol.~92, no.~4, pp. 1079--1092, 2012.

\bibitem{de2013space}
A.~L.~F. de~Almeida, G.~Favier, and L.~R. Ximenes, ``{Space-time-frequency
  {(STF)} {MIMO} communication systems with blind receiver based on a
  generalized {PARATUCK2} model},'' \emph{IEEE Trans. Signal Process.},
  vol.~61, no.~8, pp. 1895--1909, 2013.

\bibitem{Clerck_Tutorial}
H.~Li, M.~Nerini, S.~Shen, and B.~Clerckx, ``A tutorial on beyond-diagonal
  reconfigurable intelligent surfaces: Modeling, architectures, system design
  and optimization, and applications,'' \emph{IEEE Communications Surveys \&
  Tutorials}, vol.~28, pp. 4086--4126, 2026.

\bibitem{Clerckx_CAMSAP_2023}
H.~Li, Y.~Zhang, and B.~Clerckx, ``Channel estimation for beyond diagonal
  reconfigurable intelligent surfaces with group-connected architectures,'' in
  \emph{(CAMSAP)}, 2023, pp. 21--25.

\bibitem{Sokal_Asilomar_2024}
B.~Sokal, Fazal-E-Asim, A.~L.~F. de~Almeida, H.~Li, and B.~Clerckx, ``A
  decoupled channel estimation method for beyond diagonal {RIS},'' in
  \emph{2024 58th Asilomar Conference on Signals, Systems, and Computers},
  2024, pp. 1395--1399.

\bibitem{Kenneth_WCL}
K.~B.~A. Benício, A.~L.~F. de~Almeida, B.~Sokal, Fazal-E-Asim, B.~Makki, and
  G.~Fodor, ``Tensor-based channel estimation and data-aided tracking in
  {IRS}-assisted {MIMO} systems,'' \emph{IEEE Wireless Communications Letters},
  vol.~13, no.~2, pp. 333--337, 2024.

\bibitem{Andre_EURASIP_2014}
G.~Favier and A.~de~Almeida, ``Overview of constrained {PARAFAC} models,''
  \emph{EURASIP J. Adv. Signal Process}, 2014.

\bibitem{Harshman}
R.~A. Harshman, ``Foundations of the {PARAFAC} procedure: Models and conditions
  for an ``explanatory'' multi-modal factor analysis,'' \emph{UCLA Working
  Papers in Phonetics}, vol.~16, pp. 1--84, 1970.

\bibitem{Eckart_36}
C.~Eckart and G.~Young, ``The approximation of one matrix by another of lower
  rank,'' \emph{Psychometrika}, vol.~1, no.~3, pp. 211--218, Sep 1936.

\bibitem{Du2023SemiBlindRIS}
J.~Du, X.~Luo, X.~Li, M.~Zhu, K.~M. Rabie, and F.~Kara, ``Semi-blind joint
  channel estimation and symbol detection for {RIS}-empowered multiuser mmwave
  systems,'' \emph{IEEE Communications Letters}, vol.~27, no.~1, pp. 362--366,
  2023.

\bibitem{SOKAL_Elsevier}
B.~Sokal, A.~L. {de Almeida}, and M.~Haardt, ``Semi-blind receivers for {MIMO}
  multi-relaying systems via rank-one tensor approximations,'' \emph{Signal
  Processing}, vol. 166, p. 107254, 2020.

\bibitem{stegeman2007}
A.~Stegeman and N.~D. Sidiropoulos, ``On {K}ruskal’s uniqueness condition for
  the candecomp/parafac decomposition,'' \emph{Linear Algebra and its
  Applications}, vol. 420, no.~2, pp. 540 -- 552, 2007.

\bibitem{SidNway2000}
N.~D. Sidiropoulos and R.~Bro, ``On the uniqueness of multilinear decomposition
  of n-way arrays,'' \emph{Journal of Chemometrics}, vol.~14, no.~3, pp.
  229--239, 2000.

\bibitem{LDL2008}
L.~De~Lathauwer, ``Decompositions of a higher-order tensor in block
  terms—part {I}: Lemmas for partitioned matrices,'' \emph{SIAM Journal on
  Matrix Analysis and Applications}, vol.~30, no.~3, pp. 1022--1032, 2008.

\bibitem{Favier2019}
P.~M.~R. de~Oliveira, C.~A.~R. Fernandes, G.~Favier, and R.~Boyer, ``{PARATUCK}
  semi-blind receivers for relaying multi-hop {MIMO} systems,'' \emph{Digital
  Signal Processing}, vol.~92, pp. 127 -- 138, 2019.

\end{thebibliography}

\vfill\pagebreak

\end{document}